\documentclass[aps,article]{revtex4}%
\usepackage{amsfonts}
\usepackage{amsmath}
\usepackage{amssymb}
\usepackage{graphicx}%
\begin{document}
\title{A chiral model for $\overline{q}q$ and $\overline{q}\overline{q}qq$ mesons }
\author{Mauro Napsuciale}
\address{Instituto de F\'{\i}sica, Universidad de Guanajuato \\
Lomas del Bosque 103, Fracc. Lomas del Campestre, 37150, 
Le\'{o}n Gto. M\'{e}xico}
\author{Sim\'{o}n Rodr\'{\i}guez}
\address{Instituto de F\'{\i}sica, Universidad de Guanajuato \\
Lomas del Bosque 103, Fracc. Lomas del Campestre, 37150, 
Le\'{o}n Gto. M\'{e}xico}
\begin{abstract}
We point out that the spectrum of pseudoscalar and scalar mesons exhibits a
cuasi-degenerate chiral nonet in the energy region around $1.4~$GeV whose
scalar component has a slightly inverted spectrum. Based on the empirical
linear rising of the mass of a hadron with the number of constituent quarks
which yields a mass around $1.4~$GeV for tetraquarks, we conjecture that this
cuasi-chiral nonet arises from the mixing of a chiral nonet composed \ of
tetraquarks with conventional $\overline{q}q$ states. We explore this
possibility in the framework of a chiral model assuming a tetraquark chiral
nonet around $1.4~$GeV with chiral symmetry realized directly. We stress that
$U_{A}(1)$ transformations can distinguish $\overline{q}q$ from tetraquark
states, although it cannot distinguish specific dynamics in the later case. We
find that the measured spectrum is consistent with this picture. In general,
pseudoscalar states arise as mainly $\overline{q}q$ states but scalar states
turn out to be strong admixtures of $\overline{q}q$ and tetraquark states. We
work out also the model predictions for the most relevant couplings and
calculate explicitly the strong decays of the $a_{0}(1450)$ and $K_{0}^{\ast
}(1430)$ mesons. From the comparison of some of the predicted couplings with
the experimental ones we conclude that observable for the isovector and
isospinor sectors are consistently described within the model. The proper
description of couplings in the isoscalar sectors would require the
introduction of glueball fields which is an important missing piece in the
present model.

\end{abstract}
\maketitle


\section{Introduction.}

The light scalar mesons have been under intense debate during the past few
years. The fact that these states have the same quantum numbers as the vacuum
renders the understanding of their properties a primary concern since it can
shed some light on the non-trivial structure of the vacuum in QCD and
ultimately on the understanding of the mechanism for confinement in this
theory. The existence of the $f_{0}(980)$ and the $a_{0}(980)$ was firmly
established since the early 70's due to their relatively narrow width. Both
couple strongly to $\bar{K}K$ and lie so close to the $\bar{K}K$ threshold at
987 MeV, that their shapes are distorted by threshold effects in such a way
that a naive interpretation in terms of constituents must be taken with great
care. The correct interpretation of the $f_{0}$ and $a_{0}$ requires a coupled
channel scattering analysis, being $\pi\pi$ and $\bar{K}K$ for the $f_{0}$,
and $\pi\eta$ and $\bar{K}K$ for the $a_{0},$ the relevant channels. Wealth of
work has been done trying to understand this phenomena and even a strong
isospin violation has been suggested due to the strong coupling of these
states to the $\bar{K}K$\ channel \cite{Achasov,Close,Lucio}.

In addition to these states there are signals for other light scalar mesons.
In spite of the fact that a light isoscalar scalar, the so-called sigma meson,
was predicted by unitarized models \cite{TornqvistUQM, EefUMM}, and effective
models \cite{Scadron,Syracusepipi, Ishidasigma,Mauro,TornqvistEJP}, it
reentered the PDG classification of particles as the $f_{0}(400-1200)$
\cite{RPP98} only after its prediction by the unitarization of the results of
chiral perturbation theory as a state that do not survive in the large $N_{c}$
limit \cite{UCHPT}. After an intense activity on the experimental side over
the past few years, compelling evidence has been accumulated for the existence
of this state and its mass has been measured \cite{E791sigmaf0}. Nowadays we
can safely say that we have a universal acceptance for the existence of
enhancements at low energy in the s-wave isoscalar meson-meson scattering but
the interpretation of this phenomena is still controversial. Concerning the
isoscalar channel a resonance with a mass around $900$ MeV has been predicted
by unitarized quark models \cite{EefUMM}, effective models
\cite{Scadron,Mauro}, and unitarized chiral perturbation theory calculations
\cite{UCHPT}. The $\pi K$ s-wave data has been reanalyzed finding a relatively
strong attraction in the $I=1/2$ channel\ \cite{Syracusekappa, Ishidakappa}.
This enhancement in the isoscalar channel of $\pi K$\ scattering is identified
with an S-matrix pole at approximately 900 MeV ( $\kappa$(900)). Recently,
further evidence was reported for this resonance \cite{E791kappa}. In spite of
this, its existence is not generally accepted \cite{Pennington}.

As a result of different analysis a general consensus is emerging on the
existence of a light scalar nonet composed of the $\sigma(500),$ $f_{0}(980),$
$\kappa(900)$ and $a_{0}(980)$ states \cite{Scadron, Mauro,
EefUMM,Syracuseput,E791sigmaf0,E791kappa,UCHPT,Ishidakappa}. In particular, a
recent analysis of three different independent aspects of the problem: the
movement of the poles from the physical to the $SU(3)$ limit, the couplings to
two pseudoscalar physical mesons and couplings to $SU(3)$ eigenstates,
concludes that the lowest lying scalar nonet is composed of these states
\cite{Oller}.

The spectrum of this nonet contradicts the naive quark models expectations
which requires $^{3}P_{0}$ states to be heavier than the $^{3}S_{1}$ states.
Some members of the light scalar nonet have a mass smaller than the
corresponding members of the vector nonet. On the other hand, the nearly
degeneracy of the $f_{0}(980)$ and $a_{0}(980)$ suggest an $\omega-\rho$ like
quark composition but this is incompatible with the strong coupling of the
$f_{0}(980)$ to $\overline{K}K$ system. Without exhausting the list of
problems in the scalar sector we recall that the quark structure as probed by
the electromagnetic interaction is definitively not consistent with a be naive
$\overline{q}q$ composition \cite{Barnes, MauroMWPF}. The latter problem was
evident shortly after the discovery of the $f_{0}(980)$ and $a_{0}(980)$
mesons and alternative models for its internal structure were proposed. A
$\bar{q}^{2}q^{2}$ structure was suggested by Jaffe \cite{Jaffe:1977} and a
\textquotedblleft molecular\textquotedblright\ (clustered four-quark)
composition was put forth in \cite{Weinstein:1990}. The former proposal has
gained renewed interest due to a striking prediction of this model which is
consistent with the spectrum of the light scalar nonet: an \emph{inverted mass
spectrum} for four-quark states, when compared with a $\bar{q}q$ nonet, due to
the presence of a hidden $\bar{s}s$ pair in some of the $\bar{q}^{2}q^{2}$
states \cite{Jaffe:lat}. The quark content for the isotriplet and one of the
isosinglets is $\{u\bar{d}s\bar{s},(u\bar{u}-d\bar{d})s\bar{s}/\sqrt{2}%
,d\bar{u}s\bar{s}\}$ and $(u\bar{u}+d\bar{d})s\bar{s}/\sqrt{2}$, being these
the heavier states. The remaining isosinglet has a $u\bar{d}d\bar{u}$
structure without strange quarks therefore being the lighter one. The
isospinors contain a strange quark and should lie in between.

Since the mass of a hadron should increase roughly linearly with the number of
constituent quarks, four-quark states are naively expected around $1.5\ $GeV.
In Jaffe's picture, a magnetic gluon exchange interaction accounts for the
lowering of this scale down to around $900\ $MeV in such a way that the lowest
lying scalar nonet can be in principle identified as four-quark states. In
this model, the nearly degeneracy of \ the $f_{0}(980)$ and $a_{0}(980)$ and
their strong coupling to $\bar{K}K$ is just a consequence of the flavor
structure. The couplings to $\bar{q}q$ mesons follow from the flavor
structure. In general, this formalism accounts for the spectrum of the lowest
lying scalar mesons. A crucial test of the quark structure of the scalar
mesons would be the systematic study of their couplings, specially the
calculation of two photon decays of neutral scalars which to the best of our
knowledge is still lacking. The calculation of electromagnetic transitions
such as $V\rightarrow S\ \gamma$ are also necessary to understand the recently
presented data by the experimental groups at $\phi$ factories \cite{novo,kloe}
from this perspective. Finally, four-quark states are expected to mix with
conventional $\overline{q}q$ states to yield physical mesons and this mixing
should be accounted for in this formalism.

A different explanation to the light scalar spectrum was proposed in
\cite{Mauro,TornqvistEJP,MS,MWM,MauroMWPF} based on phenomenological
Lagrangians for QCD formulated firstly in \cite{schechter}. In this formalism,
the \emph{inverted mass spectrum }of the lightest scalar nonet arises from an
interplay between the spontaneous breaking of chiral symmetry (SB$\chi$S) and
a trilinear interaction between fields. This interaction is a remnant at the
hadron level of the six-quark interaction due to instantons in QCD
\cite{tHooft:ins} and is trilinear for $\overline{q}q$ states only, hence
scalars are interpreted as $\overline{q}q$ in this framework. Under SB$\chi$S,
two of the quarks entering in the six-quark 't Hooft interaction acquires
non-vanishing v.e.v. generating flavour-dependent mass terms which mix the
fields and drastically modify the mass spectrum for scalars as compared with
naive quark model expectations. This mechanism yields also an inverted mass
spectrum and describes this phenomena and the distorting of the pseudoscalar
spectrum in a unified way \cite{MS}. In this formalism couplings between
scalar and scalar mesons are dictated by chiral symmetry and can be related to
meson masses. The small two photon decay width of scalars is also
satisfactorily described in this framework \cite{two-photonsML}. Indeed, these
decays are induced by loops of charged meson in this framework, the final
photon being emitted by the charged mesons in the loop. Interestingly, the
main contribution to both $f_{0}(980)$ and $a_{0}(980)$ decays into two
photons comes from loops of charged kaons. In the former case, a dominant
contribution is naively expected to come from charged pion loops since this
meson predominantly decay into pions and the latter are lighter than kaons.
However, $f_{0}(980)$ coupling to pions is small and it is only because the
$\overline{K}K$ channel has vanishing phase space that this meson decay mainly
into two pions. The two photon decay of the $f_{0}(980)$, tests the
\emph{couplings}\textit{\ }and phase space plays no role. Therefore this decay
provides further evidence for the strong coupling of the $f_{0}(980)$ to a
pair of kaons and ultimately for its internal bare structure as predominantly
$\overline{s}s$ with some dressing of $\overline{K}K$ due to quantum
fluctuations. A similar effect is seen in the rare $\phi\rightarrow\pi^{0}%
\pi^{0}\gamma$ decay but in this case\ for the sigma meson which was expected
to dominate the low energy region of the $\pi^{0}\pi^{0}$\ invariant mass.
This region was accurately measured by the KLOE Coll. \cite{kloe} finding no
\textquotedblleft bump\textquotedblright\ associated to the sigma. Instead,
this Coll. finds a strange pattern which is attributed to a $\sigma-f_{0}$
interference. In our view, this is another beautiful manifestation of the
mechanism behind the inverted mass spectrum of the light scalar mesons.
Indeed, in order to the sigma manifest in the dipion invariant mass spectrum
it is necessary that this meson couple not only to the \textit{final } state
but also to the \textit{production} mechanism. Since the involved particles
have no electric charge, in addition to mechanisms involving vector mesons
which are under control, this decay proceeds through the chain $\phi
\rightarrow S\gamma\rightarrow\pi^{0}\pi^{0}\gamma$ with the first decay
proceeding via loops of charged mesons. The decaying $\phi$ is an
$\overline{s}s$ state and does not couple to pions, thus the leading
contributions come from kaon loops. The $f_{0}$ strongly couple to kaons and
although weakly to pions it is sufficient to show up in the dipion mass
spectrum. As for the sigma meson, it would dominate the dipion spectrum at low
energies if its coupling to kaons were large enough. However, in the model
this coupling goes as $(m_{\sigma}^{2}-m_{K}^{2})/\Lambda$ with $\Lambda$ a
typical hadronic scale, thus being small for a light sigma with a mass around
the kaon mass. The sigma contribution in the low energy dipion mass region
turns out to be of the order of the $f_{0}$ contribution yielding the observed
interference \cite{bar-gua}. Both, the interference pattern in $\phi
\rightarrow\pi^{0}\pi^{0}\gamma$\ decay and the two photon decay of the
\ $f_{0}(980)$ are sensitive to the scalar mixing angle. The proper
description of these processes is consistent with a mixing angle $\varphi
_{s}\approx-15^{o}$. This angle is consistent with the spectrum of the light
scalars in the model \cite{Mauro, MS}, and a similar value is obtained in the
independent analysis of the properties of light scalar mesons presented in
\cite{Oller}. In summary, \emph{the fundamental reason why the sigma does not
show up clearly in the dipion mass spectrum is that it does not couple to the
production mechanism, a fact related to the mechanism for the inverse mass
spectrum!}

Some other $\phi$ radiative decays have been studied in this framework
\cite{bar-gua} yielding results which accommodate the experimental data from
high luminosity $\phi$ factories, thus rendering a simple framework where the
measured properties of scalar are satisfactorily explained. It must be
mentioned however that the model is still too crude and there are some aspects
which do not work that well. In particular, the $\eta-\eta^{\prime}$ mixing
angle as estimated using the pseudoscalar spectrum as input is too small:
$-5{{}^{\circ}}\leq\theta_{p}\leq2{{}^{\circ}}$ in the singlet-octet basis or
$49{{}^{\circ}}\leq\varphi_{p}\leq57{{}^{\circ}}$ in the strange-nonstrange
basis \cite{MS,MWM}. An estimation of this angle from the two gamma decays of
$\eta$ and $\eta^{\prime}$ using the strong contribution to the singlet
anomalies as predicted by the model, the electromagnetic contributions to such
anomalies as calculated in QCD, and the anomaly matching arguments by 't Hooft
yield $39{{}^{\circ}}\leq\varphi_{p}\leq41{{}^{\circ}}$ ($-15{{}^{\circ}}%
\leq\theta_{p}\leq-13{{}^{\circ}}$ in the singlet-octet basis) \cite{MS1}.

The three flavor linear sigma model was also used in \cite{syracuse} as a
\textquotedblleft toy model\textquotedblright\ to study properties of light
mesons. A convenient unitarization procedure was proven to give encouraging
results for meson-meson scattering. In addition, the possible quark
composition of the light scalar (and pseudoscalar) fields was discussed on the
light of the chiral transformation properties of the corresponding quark
structures. The conclusion is that non-standard structures such as
$\overline{q}\overline{q}qq$ have the same transformation properties under
chiral rotations as the conventional $\overline{q}q$ structures and they can
not be distinguished by chiral effective Lagrangians. Indeed, the conventional
mapping between meson fields and quark structures considers the matrix
\begin{equation}
M_{a}^{b}\sim\overline{{q}}_{bA}^{R}\ q_{aA}^{L},\label{qb-q}%
\end{equation}
where $a$ ( $A$ ) is a flavor (color) index, and $q_{L}=\frac{1}{2}\left(
1+\gamma_{5}\right)  q$ , $q_{R}=\frac{1}{2}\left(  1-\gamma_{5}\right)  q$
stand for the left- and right-handed quark projections respectively, as
realizing a $q\bar{q}$ composite color-singlet meson field. The transformation
properties of meson fields under chiral rotations on $q_{L}$ and $q_{R}$ are
induced from this mapping. However, there exists some other possible
structures such as the \textquotedblleft molecule\textquotedblright\ field
\begin{equation}
\mathcal{M}_{a}^{\ \ b}=\epsilon_{acd}\epsilon^{bef}(M{^{\dagger})}%
_{e}^{\ \ c}(M{^{\dagger})}_{f}^{\ \ d},\label{molecule}%
\end{equation}
or the four-quark composition in a $\ $diquark-anti-diquark ( $qq\bar{q}%
\bar{q}$ ) configuration, namely%
\[
D^{gf}=(D{_{L}^{gA})}^{\dagger}D_{R}^{fA},
\]
with the diquark fields $D_{L},D_{R}$ in a flavor-triplet and color-triplet
state given by
\begin{equation}
D_{L}^{gA}=\epsilon^{gab}\epsilon^{ABC}q_{aB}^{T}C^{-1}q_{bC}^{L}\ ,\quad
D_{R}^{gA}=\epsilon^{gab}\epsilon^{ABC}q_{aB}^{T}C^{-1}q_{bC}^{R}%
,\label{di-quark}%
\end{equation}
where $C$ denotes the charge conjugation matrix. In addition, it is also
possible to have more complicated structures such as the diquark- diquark
composite with the diquarks in a flavor-triplet, color-sextet configuration.
Under chiral $SU(3)_{L}\times SU(3)_{R}$ transformations these fields
transform as
\begin{equation}
F\rightarrow U_{L}FU_{R}^{\dagger}%
\end{equation}
where $F$ denotes $M,$ $\mathcal{M}$ or $D,$ and $U_{L}$ and $U_{R}$ are
independent unitary, unimodular matrices associated with the transformations
on $q_{L}$ and $q_{R}.$ These meson fields have also identical transformation
properties under parity and charge conjugation:
\begin{equation}
C:\quad F\rightarrow F^{T},\quad\quad P:\quad F(\mathbf{x})\rightarrow
F^{\dagger}(-\mathbf{x}).
\end{equation}
However, under $U(1)_{A}$ transformations $q_{a}^{L}\rightarrow e^{i\theta
}q_{a}^{L}$, $q_{a}^{R}\rightarrow e^{-i\theta}q_{a}^{R}$ quark-antiquark and
four-quark structures transform differently
\begin{equation}
M\rightarrow e^{2i\theta}M,\quad\mathcal{M}\rightarrow e^{-4i\theta
}\mathcal{M},\quad D\rightarrow e^{-4i\theta}D.\label{M1U1A}%
\end{equation}

Since the chiral $SU(3)_{L}\times SU(3)_{R}$ transformation properties of all
these fields are identical, in \cite{syracuse} the conclusion is reached that
we cannot assign a specific quark structure to the fields used in the
construction of phenomenological chiral Lagrangians realizing this symmetry,
which on the other side are the appropriate tools for the long distance
description of the physics of mesons composed by light quarks, whatever this
composition be. However, the full chiral symmetry at the level of massless QCD
is actually $U(3)_{L}\times U(3)_{R}$ and the violation of $U(1)_{A}$ by
non-perturbative effects can distinguish at least a $\overline{q}q$ state from
a four-quark state in any of its internal configurations. This is clear if we
consider the effects of instantons which at the quark level\ generate an
effective six-quark interaction manifesting in a three-meson determinantal
interaction at the hadron level which is claimed in \cite{Mauro, MS,MWM} to be
responsible for the inverse scalar spectrum.

The existence of two scalar nonets, one below $1\ $GeV and another one around
$1.4\ $GeV, lead to the exploration of two nonet models. A first step in this
direction was given in \cite{twononetsyr} where a light four-quark scalar
nonet and a heavy $\overline{q}q$ -like nonet are mixed up to yield physical
scalar mesons above and below $1$ GeV. In the same spirit, in Ref.
\cite{syracuse} a linear sigma model is coupled to a nonet field with well
defined transformation properties under chiral rotations and trivial dynamics,
except for a $U_{A}(1)$ (non-determinantal) violating interaction. The
possibility for the pseudoscalar light mesons to have a small four-quark
content was speculated on the light of this \textquotedblleft
toy\textquotedblright\ model. An alternative approach was formulated in
\cite{TC} where two linear sigma models were coupled using the very same
interaction as in \cite{twononetsyr}. The novelty here is a Higss mechanism at
the hadron level in such a way that the members of a pseudoscalar nonet are
\textquotedblleft eaten up\textquotedblright\ by axial vector mesons and one
ends up with only one nonet of pseudoscalars and two nonets of scalars. The
model is used as a specific realization of the conjectured \textquotedblleft
energy-dependent\textquotedblright\ composition of scalar mesons based on a
detailed analysis of the whole experimental situation \cite{TC}.

Beyond the lowest lying $\sigma(500),$ $f_{0}(980),$ $\kappa(900)$ and
$a_{0}(980)$ states, the next group of scalars listed by Particle Data Group
are all in the energy region around $1.4\ $GeV: $f_{0}(1370),$ $K_{0}^{\ast
}(1430),$ $a_{0}(1450),$ and $f_{0}(1500)$ \cite{PDG}.\ In addition, we have
the $f_{0}(1710)$ at a slightly higher mass \cite{PDG}. The main point of this
paper is to note that if we consider the scalar states around $1.4\ $GeV as
the members of a nonet, then it presents a slightly inverted mass spectrum:
the heavier states are the cuasi-degenerate isotriplet and isosinglet states
$a_{0}(1450),$ and $f_{0}(1500)$, the lightest one is the isosinglet
$f_{0}(1370)$ and the isovector $K_{0}^{\ast}(1430)$ is in between. This
structure is characteristic of a four-quark nonet. Furthermore, a look onto
the pseudoscalar side at the same energy yields the following states:
$\eta(1295),\pi(1300),$ $\eta(1450),$ $K(1469)$ \cite{PDG}. Thus \emph{data
seems to indicate the existence of a cuasi-degenerate chiral nonet around
}$\emph{1.4}$\emph{\ GeV} \emph{whose scalar component has a slightly inverted
mass spectrum}. On the other hand, the linear rising of the mass of a hadron
with the number of constituent quarks indicates that four-quark states should
lie slightly below $1.5$ GeV. This lead us to conjecture that this chiral
nonet comes from tetraquark states mixed with conventional $\overline{q}q$
mesons to form physical mesons. The cuasi-degeneracy of this chiral nonet
indicates that chiral symmetry is realized in a direct way for tetraquark states.

In this work we report results on the implementation of this idea in the
framework of an effective chiral Lagrangian. We start with two chiral nonets,
one around $1.4$ GeV with chiral symmetry realized directly, and another one
at low energy with chiral symmetry spontaneously broken. In contrast to
previous studies, mesons in the \textquotedblleft heavy\textquotedblright%
\ nonet are considered as four-quark states. The nature of these states is
distinguished from conventional $\overline{q}q$ mesons by terms breaking the
$U_{A}(1)$ symmetry. We introduce also mass terms appropriate to four-quark
states which yield an inverted spectrum for the \textquotedblleft
pure\textquotedblright\ four-quark structured fields. These states mix with
conventional $\overline{q}q$ states. We fix the parameters of the model using
as input the masses for $\pi(137)$, $a_{0}(980)$, $K(495),$ $\eta(548),$
$\eta^{\prime}(957),$ $\eta(1295)$ in addition to the weak decay constants
$f_{\pi}$ and $f_{K}$. The model gives definite predictions for the masses of
the remaining members of the nonets. Couplings of all mesons entering the
model are also predicted. The outcome of the fit allow us to identify the
remaining mesons as the $\pi(1300),\eta(1450),$ $K(1469)$ on the pseudoscalar
side and $f_{0}(1370),$ $K_{0}^{\ast}(1430),$ $a_{0}(1450),$ $f_{0}(1500)$ on
the scalar side. In general, isospinor and isovector pseudoscalar mesons arise
as weak mixing of $\overline{q}q$ and tetraquark fields. In contrast, scalars
in these isotopic sectors are strongly mixed. Concerning the isoscalar sector
we obtain a strong mixing between $\overline{q}q$ and tetraquarks for both
scalar and pseudoscalar fields. We remark that results for isosinglet scalars
are expected to get modified by the mixing of pure $\overline{q}q$ and
four-quark- structured mesons with the lowest lying scalar glueball field. We
work out also the model predictions for the most relevant couplings and
calculate explicitly the strong decays of the $a_{0}(1450)$ and $K_{0}^{\ast
}(1430)$ mesons.

\section{The model.}

We start by fixing our conventions and notations. The effective model is
written in terms of \ \textquotedblleft standard\textquotedblright\ (
$\overline{q}q$-like ground states) meson fields and \ \textquotedblleft
non-standard\textquotedblright( four-quark like states) fields denoted
as$\ \Phi=S+iP,$ $\hat{\Phi}=\hat{S}+i\hat{P}$ respectively. Here, $S$ ,
$\hat{S} $ and $P$ , $\hat{P}$ denote matrix fields defined as%
\begin{equation}
F\equiv\frac{1}{\sqrt{2}}f_{i}\lambda_{i}\qquad i=1,...,7,ns,s\label{notation}%
\end{equation}
where $f_{i}$ \ stands for a generic field, $\lambda_{i}$ $(i=1,...,7)$ denote
Gell-Mann \ matrices and we work in the strange-non-strange basis for the
isoscalar sector, i.e. we use $\lambda_{ns}=diag(1,1,0)$ \ and $\lambda
_{s}=\sqrt{2}diag(0,0,1)$. Explicitly, the bare $S$, $\hat{S}$ scalar and $P$,
$\hat{P}$ pseudoscalar nonets are given by%
\begin{equation}
S=\left(
\begin{array}
[c]{ccc}%
\frac{S_{ns}+S^{0}}{\sqrt{2}} & S^{+} & Y^{+}\\
S^{-} & \frac{S_{ns}-S^{0}}{\sqrt{2}} & Y^{0}\\
Y^{-} & \bar{Y}^{0} & S_{s}%
\end{array}
\right)  ;P=\left(
\begin{array}
[c]{ccc}%
\frac{H_{ns}+p^{0}}{\sqrt{2}} & p^{+} & X^{+}\\
p^{-} & \frac{H_{ns}-p^{0}}{\sqrt{2}} & X^{0}\\
X^{-} & \bar{X}^{0} & H_{s}%
\end{array}
\right)  ,
\end{equation}%
\begin{equation}
\hat{S}=\left(
\begin{array}
[c]{ccc}%
\frac{\hat{S}_{ns}+\hat{S}^{0}}{\sqrt{2}} & \hat{S}^{+} & \hat{Y}^{+}\\
\hat{S}^{-} & \frac{\hat{S}_{ns}-\hat{S}^{0}}{\sqrt{2}} & \hat{Y}^{0}\\
\hat{Y}^{-} & \overline{\hat{Y}}^{0} & \hat{S}_{s}%
\end{array}
\right)  ;\hat{P}=\left(
\begin{array}
[c]{ccc}%
\frac{\hat{H}_{ns}+\hat{p}^{0}}{\sqrt{2}} & \hat{p}^{+} & \hat{X}^{+}\\
\hat{p}^{-} & \frac{\hat{H}_{ns}-\hat{p}^{0}}{\sqrt{2}} & \hat{X}^{0}\\
\hat{X}^{-} & \overline{\hat{X}}^{0} & \hat{H}_{s}%
\end{array}
\right)  .\label{4qfields}%
\end{equation}
Next, we realize the idea of a chiral nonet around 1.4 GeV using an effective
Lagrangian written in terms of four-quark structured fields, $\hat{\Phi}$,
with chiral symmetry realized linearly and directly ( $\hat{\mu}^{2}>0$)%
\begin{align}
\mathcal{L}(\hat{\Phi}) &  =\mathcal{L}_{sym}(\hat{\Phi})+\mathcal{L}%
_{SB}(\hat{\Phi}),\label{L4q}\\
\mathcal{L}_{sym}(\hat{\Phi}) &  =\frac{1}{2}\left\langle \partial_{\mu}%
\hat{\Phi}\text{ }\partial^{\mu}\hat{\Phi}^{\dagger}\right\rangle -\frac
{\hat{\mu}^{2}}{2}\left\langle \hat{\Phi}\hat{\Phi}^{\dagger}\right\rangle
-\frac{\widehat{\lambda}}{4}\left\langle (\hat{\Phi}\hat{\Phi}^{\dagger}%
)^{2}\right\rangle ,\nonumber\\
\mathcal{L}_{SB}(\hat{\Phi}) &  =-\frac{\hat{c}}{4}\left\langle \mathcal{M}%
_{Q}(\hat{\Phi}\hat{\Phi}^{\dagger}+\hat{\Phi}^{\dagger}\hat{\Phi
})\right\rangle .\nonumber
\end{align}
Here, $\hat{\mu}$ sets the scale at which four-quark states lie. This scale is
expected to be slightly below 1.4 GeV. Although some of the fields in
$\hat{\Phi}$ have the same quantum numbers as the vacuum they do not acquire
vacuum expectation values (vev%
\'{}%
s) since we require a direct realization of chiral symmetry. The symmetry
breaking term in Eq. (\ref{L4q}) requires some explanation. This is an
explicit breaking term quadratic in the fields and proportional to a quadratic
quark mass matrix. The point is that in a quiral expansion the quark mass
matrix has a non-trivial flavor structure and enters as an external scalar
field. In the case of a four-quark nonet there must be breaking terms with the
appropriate flavor structure. This matrix is constructed according to$\ $Eq.
(\ref{molecule}) as%
\begin{equation}
(\mathcal{M}_{Q})_{a}^{\ \ b}=\frac{1}{2}\epsilon_{acd}\epsilon^{bef}%
(\mathcal{M}_{q}{^{\dagger})}_{e}^{\ \ c}(\mathcal{M}_{q}{^{\dagger})}%
_{f}^{\ \ d},\label{M4q}%
\end{equation}
where $\mathcal{M}_{q}=Diag(m,m,m_{s})$ stands for the conventional quark mass
matrix in the good isospin limit which we will consider henceforth.
Explicitly, we obtain $\mathcal{M}_{Q}=Diag(mm_{s},mm_{s},m^{2})$. This
structure yields to pure 4q-structured fields an inverted spectrum with
respect to conventional states. A word of caution is necessary concerning the
notation in Eq. (\ref{4qfields}). The matrix field for four-quark states has a
schematic quark content%
\[
\hat{\Phi}\sim\left(
\begin{array}
[c]{ccc}%
\overline{q}q\overline{s}s & \overline{q}q\overline{s}s & \overline
{q}q\overline{q}s\\
\overline{q}q\overline{s}s & \overline{q}q\overline{s}s & \overline
{q}q\overline{q}s\\
\overline{q}q\overline{q}s & \overline{q}q\overline{q}s & \overline
{q}q\overline{q}q
\end{array}
\right)  ,
\]
where $q$ denote a $u$ or $d$ quark. The subindex $s$ and $ns$ in these fields
refer to the notation for $SU(3)$ matrices in Eq. (\ref{notation}) but do not
corresponds with the hidden quark-antiquark content, e.g. $\hat{S}_{ns}%
\sim\overline{q}q\overline{s}s$ and $\hat{S}_{s}\sim\overline{q}q\overline
{q}q$.

Conventional $\overline{q}q$-structured fields are introduced in a chirally
symmetric way with chiral symmetry spontaneously broken. We introduce also an
instanton inspired breaking for the $U_{A}(1)$ symmetry. Notice that the
determinantal interaction is appropriate for $\overline{q}q$-structured fields
but not for four-quark fields since this is a six-quark interaction%
\begin{align}
\mathcal{L}(\Phi) &  =\mathcal{L}_{sym}(\Phi)+\mathcal{L}_{A}+\mathcal{L}%
_{SB}(\Phi),\label{L2q}\\
\mathcal{L}_{sym}(\Phi) &  =\frac{1}{2}\left\langle \partial_{\mu}\Phi
\partial^{\mu}\Phi^{\dagger}\right\rangle -\frac{\mu^{2}}{2}\left\langle
\Phi\Phi^{\dagger}\right\rangle -\frac{\lambda}{4}\left\langle (\Phi
\Phi^{\dagger})^{2}\right\rangle ,\nonumber\\
\mathcal{L}_{A} &  =-B\left(  \det\Phi+\det\Phi^{\dagger}\right)
,\quad\mathcal{L}_{SB}(\Phi)=\frac{b_{0}}{\sqrt{2}}\left\langle \mathcal{M}%
_{q}\left(  \Phi+\Phi^{\dagger}\right)  \right\rangle .\label{LSB1}%
\end{align}
The Lagrangian in Eqs. (\ref{L2q}, \ref{LSB1}) is the model used in
\cite{Mauro,TornqvistEJP,MS} and in the second of Ref. \cite{tHooft:ins},
except for an OZI rule violating term $-\frac{\lambda^{\prime}}{4}\left\langle
\Phi\Phi^{\dagger}\right\rangle ^{2}$. When this term is included in the
present context the coupling $\lambda^{\prime}$ turns out to be consistent
with zero. The Lagrangian
\[
\mathcal{L}_{sym}(\Phi,\hat{\Phi})=\mathcal{L}_{sym}(\Phi)+\mathcal{L}%
_{sym}(\hat{\Phi})
\]
is invariant under the independent chiral transformations%
\begin{equation}
\Phi\rightarrow U_{L}(\alpha_{L})\Phi U_{R}^{\dagger}(\alpha_{R}),\qquad
\hat{\Phi}\rightarrow\hat{\Phi};\qquad\qquad\Phi\rightarrow\Phi,\qquad
\hat{\Phi}\rightarrow\hat{U}_{L}(\hat{\alpha}_{L})\hat{\Phi}\hat{U}%
_{R}^{\dagger}(\hat{\alpha}_{R}),
\end{equation}
i.e. it has $\left(  U(3)\times U(3)\right)  ^{2}$ symmetry. This symmetry can
be explicitly broken down to $SU(3)\times U(3)$ by the interaction
\begin{equation}
\mathcal{L}_{\epsilon^{2}}=-\frac{\epsilon^{2}}{4}\left\langle \Phi\hat{\Phi
}^{\dagger}+\hat{\Phi}\Phi^{\dagger}\right\rangle .\label{Leps}%
\end{equation}
The anomaly term and quark mass terms in Eqs. (\ref{LSB1}, \ref{L4q}) break
the latter symmetry down to isospin. Finally, since we are considering quark
mass matrix entering as an external scalar field, we also consider the
following mass terms%
\begin{equation}
\mathcal{L}_{SB}=\frac{\hat{b}_{0}}{\sqrt{2}}\left\langle \mathcal{M}_{q}%
(\hat{\Phi}+\hat{\Phi}^{\dagger})\right\rangle +\frac{\hat{d}}{\sqrt{2}%
}\left\langle \mathcal{M}_{Q}(\hat{\Phi}+\hat{\Phi}^{\dagger})\right\rangle .
\end{equation}
The linear terms in (\ref{LSB1}) induce scalar-to-vacuum transitions which
instabilize the vacuum. We shift to the true vacuum, $S\rightarrow S-V$, where
$V$ stands for the vacuum expectation value of $S$, $V=diagg(a,a,b).$ This
mechanisms generates new mass terms and interactions which we organize as%
\begin{equation}
\mathcal{L}=\sum_{n}\mathcal{L}_{\left(  n\right)  }%
\end{equation}
where $\mathcal{L}_{\left(  n\right)  }$ collects terms of order $\phi^{n}$.
These terms are explicitly given by%
\begin{equation}
\mathcal{L}_{\left(  0\right)  }=-\frac{\mu^{2}}{2}(2a^{2}+b^{2}%
)-\frac{\lambda}{4}(2a^{4}+b^{4})-2Ba^{2}b+\frac{b_{0}}{\sqrt{2}}\left\langle
\mathcal{M}_{q}V\right\rangle ,\label{L0}%
\end{equation}%
\begin{align}
\mathcal{L}_{\left(  1\right)  }= &  -\mu^{2}\left\langle SV\right\rangle
-\lambda(a^{2}+ab+b^{2})\left\langle SV\right\rangle -ab(a+b)\left\langle
S\right\rangle -2B\left(  a(a+b)\left\langle S\right\rangle -a\left\langle
SV\right\rangle \right) \label{L1}\\
&  +\sqrt{2}\left\langle \mathcal{M}_{q}\left[  b_{0}S+\hat{b}_{0}\hat
{S}\right]  +\mathcal{M}_{Q}\hat{d}\hat{S}\right\rangle -\frac{\epsilon^{2}%
}{2}\left\langle \hat{S}V\right\rangle ,\nonumber
\end{align}%
\begin{align}
\mathcal{L}_{\left(  2\right)  }= &  -\frac{\mu^{2}}{2}\left\langle
S^{2}+P^{2}\right\rangle -\lambda\left\langle \left(  \left(  a+b\right)
V-ab\right)  \left(  S^{2}+P^{2}\right)  +\frac{1}{2}\left(  VS\right)
^{2}-\frac{1}{2}\left(  VP\right)  ^{2}\right\rangle ,\label{L2}\\
&  -B\left(  2a^{2}+b^{2}\right)  \left(  \left\langle S\right\rangle
^{2}-\left\langle P\right\rangle ^{2}-\left\langle S^{2}-P^{2}\right\rangle
\right)  -2\left\langle S\right\rangle \left\langle SV\right\rangle
+2\left\langle P\right\rangle \left\langle PV\right\rangle +2\left\langle
V\left(  S^{2}-P^{2}\right)  \right\rangle -\frac{\epsilon^{2}}{2}\left\langle
S\hat{S}+P\hat{P}\right\rangle \nonumber\\
&  -\frac{\hat{\mu}^{2}}{2}\left\langle \hat{S}^{2}+\hat{P}^{2}\right\rangle
-\frac{\hat{c}}{2}\left\langle \mathcal{M}_{Q}\left(  \hat{S}^{2}+\hat{P}%
^{2}\right)  \right\rangle ,\nonumber
\end{align}%
\begin{equation}
\mathcal{L}_{\left(  3\right)  }=-BZ-\lambda\left\langle V\left(  S^{3}%
+P^{2}S+SP^{2}-PSP\right)  \right\rangle ,\label{L3}%
\end{equation}%
\begin{equation}
\mathcal{L}_{\left(  4\right)  }=-\frac{\lambda}{4}\left\langle (\Phi
\Phi^{\dagger})^{2}\right\rangle -\frac{\hat{\lambda}}{4}\left\langle
(\hat{\Phi}\hat{\Phi}^{\dagger})^{2}\right\rangle .\label{L4}%
\end{equation}

Stability of the vacuum requires $\mathcal{L}_{\left(  1\right)  }=0$ which
yields%
\begin{equation}%
\begin{array}
[c]{l}%
2mb_{0}=\sqrt{2}a(\mu^{2}+2Bb+\lambda a^{2});\qquad\sqrt{2}m_{s}b_{0}=\mu
^{2}b+2Ba^{2}+\lambda b^{3},\\
2m\hat{b}_{0}+2mm_{s}\hat{d}=\frac{\epsilon^{2}a}{\sqrt{2}};\qquad\qquad
\qquad\sqrt{2}m_{s}\hat{b}_{0}+\sqrt{2}m^{2}\hat{d}=\frac{\epsilon^{2}b}{2}.
\end{array}
\label{L10}%
\end{equation}
Equations (\ref{L10}) can be rewritten in a more convenient way for future
manipulations as%
\begin{equation}%
\begin{array}
[c]{l}%
2mb_{0}=\sqrt{2}(\mu^{2}+2Bb+\lambda a^{2})a;\qquad\left(  m+m_{s}\right)
b_{0}=\frac{a+b}{\sqrt{2}}\left(  \mu^{2}+2Ba+\lambda\left(  a^{2}%
-ab+b^{2}\right)  \right)  ,\\
2m\hat{b}_{0}+2mm_{s}\hat{d}=\frac{\epsilon^{2}a}{\sqrt{2}};\qquad\qquad
\qquad\left(  m+m_{s}\right)  \hat{b}_{0}+m(m+m_{s})\hat{d}=\frac{\epsilon
^{2}}{2\sqrt{2}}\left(  a+b\right)  .
\end{array}
\label{Cero}%
\end{equation}

\section{Meson masses.}

\subsection{Isovector sector.}

All information on meson masses is contained in $\mathcal{L}_{\left(
2\right)  }$ in Eq. (\ref{L2}) which needs to be diagonalized. For the
$I\neq0$ sectors,\ the only mixing term arises from the interaction term
$\mathcal{L}_{\epsilon^{2}}$. The mass Lagrangian for isotriplet pseudoscalars
is given by%
\begin{equation}
\mathcal{L}_{\pi}=-\frac{1}{2}\left\langle \vec{P}\right\vert M_{P}\left\vert
\vec{P}\right\rangle \label{l2pi}%
\end{equation}
where%
\[
\left\vert \vec{P}\right\rangle =\left(
\begin{array}
[c]{c}%
\overrightarrow{p}\\
\overrightarrow{\hat{p}}%
\end{array}
\right)  \quad M_{P}=\left(
\begin{array}
[c]{cc}%
m_{p}^{2} & \frac{\epsilon^{2}}{2}\\
\frac{\epsilon^{2}}{2} & \hat{\mu}_{1}^{2}%
\end{array}
\right)
\]
with
\begin{equation}
m_{p}^{2}=\mu^{2}+2Bb+\lambda a^{2},\qquad\hat{\mu}_{1}^{2}=\hat{\mu}^{2}%
+\hat{c}mm_{s}.\label{mp}%
\end{equation}
Similarly, the mass term for isovector scalars turns out to be%
\[
\mathcal{L}_{a}=-\frac{1}{2}\left\langle \vec{S}\right\vert M_{S}\left\vert
\vec{S}\right\rangle
\]
where%
\begin{equation}
\left\vert \vec{S}\right\rangle =\left(
\begin{array}
[c]{c}%
\overrightarrow{S}\\
\overrightarrow{\hat{S}}%
\end{array}
\right)  \quad M_{S}=\left(
\begin{array}
[c]{cc}%
m_{S}^{2} & \frac{\epsilon^{2}}{2}\\
\frac{\epsilon^{2}}{2} & \hat{\mu}_{1}^{2}%
\end{array}
\right)  \label{isovsc}%
\end{equation}
with
\begin{equation}
m_{S}^{2}=\mu^{2}-2Bb+3\lambda a^{2}.\label{msv}%
\end{equation}
We define the diagonal pseudoscalar fields by%
\[
\left(
\begin{array}
[c]{c}%
\overrightarrow{\pi}\\
\overrightarrow{\hat{\pi}}%
\end{array}
\right)  =R(\theta_{1})\left(
\begin{array}
[c]{c}%
\overrightarrow{p}\\
\overrightarrow{\hat{p}}%
\end{array}
\right)  \equiv\left(
\begin{array}
[c]{cc}%
\cos(\theta_{1}) & -\sin(\theta_{1})\\
\sin(\theta_{1}) & \cos(\theta_{1})
\end{array}
\right)  \left(
\begin{array}
[c]{c}%
\overrightarrow{p}\\
\overrightarrow{\hat{p}}%
\end{array}
\right)  .
\]
The diagonal masses for these fields are given by%
\begin{equation}
m_{\pi}^{2}=\frac{1}{2}\left\{  m_{p}^{2}+\hat{\mu}_{1}^{2}-\sqrt{\left(
m_{p}^{2}-\hat{\mu}_{1}^{2}\right)  ^{2}+\epsilon^{4}}\right\}  ,\qquad
m_{\hat{\pi}}^{2}=\frac{1}{2}\left\{  m_{p}^{2}+\hat{\mu}_{1}^{2}%
+\sqrt{\left(  m_{p}^{2}-\hat{\mu}_{1}^{2}\right)  ^{2}+\epsilon^{4}}\right\}
,\label{mpi}%
\end{equation}
and we get the following relations for the mixing angle $\theta_{1}$%
\begin{equation}
\cos2\theta_{1}=\frac{\hat{\mu}_{1}^{2}-m_{p}^{2}}{m_{\hat{\pi}}^{2}-m_{\pi
}^{2}},\qquad\sin2\theta_{1}=\frac{\epsilon^{2}}{m_{\hat{\pi}}^{2}-m_{\pi}%
^{2}}.\label{theta1}%
\end{equation}
In addition the diagonalization procedure yields%
\[
m_{p}^{2}=m_{\pi}^{2}\cos^{2}\theta_{1}+m_{\hat{\pi}}^{2}\sin^{2}\theta
_{1},\qquad\hat{\mu}_{1}^{2}=m_{\pi}^{2}\sin^{2}\theta_{1}+m_{\hat{\pi}}%
^{2}\cos^{2}\theta_{1}.
\]
For the isovector scalar sector we define the physical fields as $a$, $A$ and
the corresponding mixing angle is denoted by $\phi_{1}$. From Eqs.
(\ref{isovsc}, \ref{msv}), the diagonal masses are%
\begin{equation}
m_{a}^{2}=\frac{1}{2}\left\{  m_{S}^{2}+\hat{\mu}_{1}^{2}-\sqrt{\left(
m_{S}^{2}-\hat{\mu}_{1}^{2}\right)  ^{2}+\epsilon^{4}}\right\}  ,\qquad
m_{A}^{2}=\frac{1}{2}\left\{  m_{S}^{2}+\hat{\mu}_{1}^{2}+\sqrt{\left(
m_{S}^{2}-\hat{\mu}_{1}^{2}\right)  ^{2}+\epsilon^{4}}\right\}  .\label{ma}%
\end{equation}
The mixing angle $\phi_{1}$ satisfy
\begin{equation}
\cos2\phi_{1}=\frac{\hat{\mu}_{1}^{2}-m_{S}^{2}}{m_{A}^{2}-m_{a}^{2}}%
,\qquad\sin2\phi_{1}=\frac{\epsilon^{2}}{m_{A}^{2}-m_{a}^{2}}.\label{phi1}%
\end{equation}

\subsection{Isospinor sector.}

The mass Lagrangians for the isodoublet fields are extracted from (\ref{L2})
as
\[
\mathcal{L}_{K}=-\frac{1}{2}\left\langle X\right\vert M_{X}\left\vert
X\right\rangle ,\qquad\mathcal{L}_{\kappa}=-\frac{1}{2}\left\langle
Y\right\vert M_{Y}\left\vert Y\right\rangle
\]
where%
\[
\left\vert X\right\rangle =\left(
\begin{array}
[c]{c}%
X\\
\hat{X}%
\end{array}
\right)  \quad M_{X}=\left(
\begin{array}
[c]{cc}%
m_{X}^{2} & \frac{\epsilon^{2}}{2}\\
\frac{\epsilon^{2}}{2} & \hat{\mu}_{1/2}^{2}%
\end{array}
\right)  ;\qquad\qquad\left\vert Y\right\rangle =\left(
\begin{array}
[c]{c}%
Y\\
\hat{Y}%
\end{array}
\right)  \quad M_{Y}=\left(
\begin{array}
[c]{cc}%
m_{Y}^{2} & \frac{\epsilon^{2}}{2}\\
\frac{\epsilon^{2}}{2} & \hat{\mu}_{1/2}^{2}%
\end{array}
\right)  .
\]
Here, $m_{X}^{2}$, $m_{Y}^{2}$ stand for the masses of the non-diagonal
fields, given by Eq. (\ref{L2}) as
\begin{equation}
m_{X}^{2}=\mu^{2}+2Ba+\lambda\left(  a^{2}-ab+b^{2}\right)  ,\qquad m_{Y}%
^{2}=\mu^{2}-2Ba+\lambda\left(  a^{2}+ab+b^{2}\right)  ,\qquad\hat{\mu}%
_{1/2}^{2}=\hat{\mu}^{2}+\frac{\hat{c}}{2}m(m+m_{s}).
\end{equation}
We denote the physical fields as $K,$ $\hat{K}$ $(\kappa,$ $\hat{\kappa})$,
and the mixing angle as $\theta_{1/2}$ $(\phi_{1/2})$ for isodoublet
pseudoscalars (scalars). The diagonalization procedure yields the following
relations%
\begin{align}
\cos2\theta_{1/2} &  =\frac{\hat{\mu}_{1/2}^{2}-m_{X}^{2}}{m_{\hat{K}}%
^{2}-m_{K}^{2}},\qquad\sin2\theta_{1/2}=\frac{\epsilon^{2}}{m_{\hat{K}}%
^{2}-m_{K}^{2}},\label{theta2}\\
\cos2\phi_{1/2} &  =\frac{\hat{\mu}_{1/2}^{2}-m_{Y}^{2}}{m_{\hat{\kappa}}%
^{2}-m_{\kappa}^{2}},\qquad\sin2\phi_{1/2}=\frac{\epsilon^{2}}{m_{\hat{\kappa
}}^{2}-m_{\kappa}^{2}},\label{phi2}%
\end{align}
where $m_{K}^{2}$, $m_{\hat{K}}^{2}$, $m_{\kappa}^{2}$ and $m_{\hat{\kappa}%
}^{2}$ stand for the physical masses given by%
\begin{equation}
m_{K}^{2}=\frac{1}{2}\left\{  m_{X}^{2}+\hat{\mu}_{1/2}^{2}-\sqrt{\left(
m_{X}^{2}-\hat{\mu}_{1/2}^{2}\right)  ^{2}+\epsilon^{4}}\right\}  ,\quad
m_{\hat{K}}^{2}=\frac{1}{2}\left\{  m_{X}^{2}+\hat{\mu}_{1/2}^{2}%
+\sqrt{\left(  m_{X}^{2}-\hat{\mu}_{1/2}^{2}\right)  ^{2}+\epsilon^{4}%
}\right\}  ,\label{m12}%
\end{equation}%
\[
m_{\kappa}^{2}=\frac{1}{2}\left\{  m_{Y}^{2}+\hat{\mu}_{1/2}^{2}-\sqrt{\left(
m_{Y}^{2}-\hat{\mu}_{1/2}^{2}\right)  ^{2}+\epsilon^{4}}\right\}  ,\quad
m_{\hat{\kappa}}^{2}=\frac{1}{2}\left\{  m_{Y}^{2}+\hat{\mu}_{1/2}^{2}%
+\sqrt{\left(  m_{Y}^{2}-\hat{\mu}_{1/2}^{2}\right)  ^{2}+\epsilon^{4}%
}\right\}  .
\]

\subsection{Isoscalar sector.}

The isosinglet sector is more involved due to the effects coming from the
$U_{A}(1)$ anomaly which when combined with the spontaneous breaking of chiral
symmetry produces mixing among four different fields. The mass Lagrangian for
the isoscalar pseudoscalar sector is extracted from (\ref{L2}) as%
\begin{equation}
\mathcal{L}_{H}=-\frac{1}{2}\left\langle H\right\vert M_{H}\left\vert
H\right\rangle
\end{equation}
where%
\begin{equation}
\left\vert H\right\rangle =\left(
\begin{array}
[c]{c}%
H_{ns}\\
H_{s}\\
\hat{H}_{ns}\\
\hat{H}_{s}%
\end{array}
\right)  ,\quad M_{H}=\left(
\begin{array}
[c]{cccc}%
m_{H_{ns}}^{2} & m_{H_{s-ns}}^{2} & \frac{\epsilon^{2}}{2} & 0\\
m_{H_{s-ns}}^{2} & m_{H_{s}}^{2} & 0 & \frac{\epsilon^{2}}{2}\\
\frac{\epsilon^{2}}{2} & 0 & m_{\hat{H}_{ns}}^{2} & 0\\
0 & \frac{\epsilon^{2}}{2} & 0 & m_{\hat{H}_{s}}^{2}%
\end{array}
\right)  ,\label{MH}%
\end{equation}
with%
\begin{align}
m_{H_{ns}}^{2} &  =\mu^{2}-2Bb+\lambda a^{2},\qquad m_{\hat{H}_{ns}}^{2}%
=\hat{\mu}^{2}+\hat{c}mm_{s},\label{mHNS}\\
m_{H_{s}}^{2} &  =\mu^{2}+\lambda b^{2},\qquad\qquad\quad\ m_{\hat{H}_{s}}%
^{2}=\hat{\mu}^{2}+\hat{c}m^{2},\label{mHS}\\
m_{H_{s-ns}}^{2} &  =-2\sqrt{2}Ba.
\end{align}
For the isoscalar-scalar sector we obtain a similar structure:
\[
\mathcal{L}_{S_{0}}=-\frac{1}{2}\left\langle S_{0}\right\vert M_{S_{0}%
}\left\vert S_{0}\right\rangle
\]
where%
\begin{equation}
\left\vert S_{0}\right\rangle =\left(
\begin{array}
[c]{c}%
S_{ns}\\
S_{s}\\
\hat{S}_{ns}\\
\hat{S}_{s}%
\end{array}
\right)  ,\quad M_{S_{0}}=\left(
\begin{array}
[c]{cccc}%
m_{S_{ns}}^{2} & m_{S_{s-ns}}^{2} & \frac{\epsilon^{2}}{2} & 0\\
m_{S_{s-ns}}^{2} & m_{S_{s}}^{2} & 0 & \frac{\epsilon^{2}}{2}\\
\frac{\epsilon^{2}}{2} & 0 & m_{\hat{S}_{ns}}^{2} & 0\\
0 & \frac{\epsilon^{2}}{2} & 0 & m_{\hat{S}_{s}}^{2}%
\end{array}
\right)  ,\label{MS}%
\end{equation}
with%
\begin{align}
m_{S_{ns}}^{2} &  =\mu^{2}+2Bb+3\lambda a^{2},\qquad m_{\hat{S}_{ns}}^{2}%
=\hat{\mu}^{2}+\hat{c}mm_{s},\\
m_{S_{s}}^{2} &  =\mu^{2}+3\lambda b^{2},\qquad\qquad m_{\hat{S}_{s}}^{2}%
=\hat{\mu}^{2}+\hat{c}m^{2},\\
m_{S_{s-ns}}^{2} &  =2\sqrt{2}Ba.
\end{align}

\subsection{Diagonalization.}

In principle, matrices in Eqs. (\ref{MH}, \ref{MS}) are diagonalized by a
general rotation in $O(4)$ containing six independent parameters. However, the
scale $\hat{\mu}$ is expected to be large compared with quark masses and we
can write%
\begin{equation}
\left(
\begin{array}
[c]{cc}%
m_{\hat{H}_{ns}}^{2} & 0\\
0 & m_{\hat{H}_{s}}^{2}%
\end{array}
\right)  =AI+\Delta\sigma_{3}\label{Delta}%
\end{equation}
where $A=(m_{\hat{H}_{ns}}^{2}+m_{\hat{H}_{s}}^{2})/2=\hat{\mu}_{1/2}^{2}$ and
$\Delta=m(m_{s}-m)\hat{c}$. In a first approximation we can neglect the second
term. In this case, the mass matrix can be written as
\begin{equation}
M_{H}\cong\left(
\begin{array}
[c]{cc}%
\tilde{M}_{H} & \frac{\epsilon^{2}}{2}I\\
\frac{\epsilon^{2}}{2}I & \hat{\mu}_{1/2}^{2}I
\end{array}
\right)  \label{aprox}%
\end{equation}
where%
\begin{equation}
\tilde{M}_{H}=\left(
\begin{array}
[c]{cc}%
m_{H_{ns}}^{2} & m_{H_{s-ns}}^{2}\\
m_{H_{s-ns}}^{2} & m_{H_{s}}^{2}%
\end{array}
\right)
\end{equation}
and $I$ denote the $2\times2$ unit matrix. This matrix can be diagonalized as
follows: first we consider a $SO(2)$ rotation%
\begin{equation}
\tilde{R}_{1}=\left(
\begin{array}
[c]{cc}%
R(\alpha) & 0\\
0 & R(\alpha)
\end{array}
\right)  \label{R1}%
\end{equation}
with%
\[
R(\alpha)=\left(
\begin{array}
[c]{cc}%
\cos\alpha & -\sin\alpha\\
\sin a & \cos\alpha
\end{array}
\right)
\]
in such a way that $\left\vert H\right\rangle _{1}=\tilde{R}_{1}\left\vert
H\right\rangle $ and the new mass matrix reads%
\begin{equation}
M_{H_{1}}=\left(
\begin{array}
[c]{cc}%
R(\alpha)\tilde{M}_{H}R^{\dag}(\alpha) & \frac{\epsilon^{2}}{2}I\\
\frac{\epsilon^{2}}{2}I & \hat{\mu}_{1/2}^{2}I
\end{array}
\right)  .
\end{equation}
The mixing angle $\alpha$ is chosen to diagonalize $R(\alpha)\tilde{M}%
_{H}R^{\dag}(\alpha)$. This yields%
\begin{equation}
\cos2\alpha=\frac{m_{H_{ns}}^{2}-m_{H_{s}}^{2}}{m_{H}^{2}-m_{H^{\prime}}^{2}%
},\qquad\sin2\alpha=\frac{-2m_{H_{s-ns}}^{2}}{m_{H}^{2}-m_{H^{\prime}}^{2}},
\end{equation}
where $m_{H}^{2},$ $m_{H^{\prime}}^{2}$ are the eigenvalues of the submatrix
$\tilde{M}_{H}$%
\begin{align}
m_{H}^{2} &  =\frac{1}{2}\left\{  m_{H_{ns}}^{2}+m_{H_{s}}^{2}-\sqrt{\left(
m_{H_{ns}}^{2}-m_{H_{s}}^{2}\right)  ^{2}+4m_{H_{s-ns}}^{4}}\right\}
,\label{mH}\\
m_{H^{\prime}}^{2} &  =\frac{1}{2}\left\{  m_{H_{ns}}^{2}+m_{H_{s}}^{2}%
+\sqrt{\left(  m_{H_{ns}}^{2}-m_{H_{s}}^{2}\right)  ^{2}+4m_{H_{s-ns}}^{4}%
}\right\}  .\label{mHp}%
\end{align}
The new mass matrix $M_{H_{1}}$ reads%
\begin{equation}
M_{H_{1}}=\left(
\begin{array}
[c]{cccc}%
m_{H}^{2} & 0 & \frac{\epsilon^{2}}{2} & 0\\
0 & m_{H^{\prime}}^{2} & 0 & \frac{\epsilon^{2}}{2}\\
\frac{\epsilon^{2}}{2} & 0 & \hat{\mu}_{1/2}^{2} & 0\\
0 & \frac{\epsilon^{2}}{2} & 0 & \hat{\mu}_{1/2}^{2}%
\end{array}
\right)  .
\end{equation}
This matrix can be diagonalized by a rotation in $SO(2)\otimes SO(2)$%
\begin{equation}
R_{2}(\beta,\beta^{\prime})=\left(
\begin{array}
[c]{cccc}%
\cos\beta & 0 & -\sin\beta & 0\\
0 & \cos\beta^{\prime} & 0 & -\sin\beta^{\prime}\\
\sin\beta & 0 & \cos\beta & 0\\
0 & \sin\beta^{\prime} & 0 & \cos\beta^{\prime}%
\end{array}
\right)  .\label{R2}%
\end{equation}
The mixing angles are given by%
\begin{align}
\cos2\beta &  =\frac{m_{H}^{2}-\hat{\mu}_{1/2}^{2}}{m_{\eta}^{2}-m_{\hat{\eta
}}^{2}},\qquad\sin2\beta=\frac{\epsilon^{2}}{m_{\hat{\eta}}^{2}-m_{\eta}^{2}%
},\\
\cos2\beta^{\prime} &  =\frac{m_{H^{\prime}}^{2}-\hat{\mu}_{1/2}^{2}}%
{m_{\eta^{\prime}}^{2}-m_{\hat{\eta}^{\prime}}^{2}},\qquad\sin2\beta
=\frac{\epsilon^{2}}{m_{\hat{\eta}^{\prime}}^{2}-m_{\eta^{\prime}}^{2}},
\end{align}
where $m_{\eta}^{2},$ $m_{\eta^{\prime}}^{2},$ $m_{\hat{\eta}}^{2}$ and
$m_{\hat{\eta}^{\prime}}^{2}$ denote the physical meson masses%
\begin{align}
m_{\eta(\hat{\eta})}^{2} &  =\frac{1}{2}\left\{  m_{H}^{2}+\hat{\mu}_{1/2}%
^{2}-(+)\sqrt{\left(  m_{H}^{2}-\hat{\mu}_{1/2}^{2}\right)  ^{2}+\epsilon^{4}%
}\right\}  ,\label{meta}\\
m_{\eta^{\prime}(\hat{\eta}^{\prime})}^{2} &  =\frac{1}{2}\left\{
m_{H^{\prime}}^{2}+\hat{\mu}_{1/2}^{2}-(+)\sqrt{\left(  m_{H^{\prime}}%
^{2}-\hat{\mu}_{1/2}^{2}\right)  ^{2}+\epsilon^{4}}\right\}  .\label{metap}%
\end{align}
In terms of the physical masses the parameters $\hat{\mu}_{1/2}^{2}$ and
$\epsilon^{4}$ are given by%
\begin{equation}
\hat{\mu}_{1/2}^{2}=\frac{m_{\eta^{\prime}}^{2}m_{\hat{\eta}^{\prime}}%
^{2}-m_{\eta}^{2}m_{\hat{\eta}}^{2}}{m_{\eta^{^{\prime}}}^{2}+m_{\hat{\eta
}^{\prime}}^{2}-m_{\eta}^{2}-m_{\hat{\eta}}^{2}},\label{mug}%
\end{equation}%
\begin{equation}
\epsilon^{4}=4\left(  \hat{\mu}_{1/2}^{2}-m_{\eta}^{2}\right)  \left(
m_{\hat{\eta}}^{2}-\hat{\mu}_{1/2}^{2}\right)  =4\left(  \hat{\mu}_{1/2}%
^{2}-m_{\eta^{^{\prime}}}^{2}\right)  \left(  m_{\hat{\eta}^{^{\prime}}}%
^{2}-\hat{\mu}_{1/2}^{2}\right)  .\label{eps}%
\end{equation}
From Eqs. (\ref{meta}, \ref{metap}), and using Eqs. (\ref{mH}, \ref{mHp}) and
$m_{H_{s-ns}}^{2}$ we get%
\begin{equation}
32B^{2}a^{2}+(m_{H_{ns}}^{2}-m_{H_{s}}^{2})^{2}=\left(  m_{\eta}^{2}%
+m_{\hat{\eta}}^{2}-m_{\eta^{\prime}}^{2}-m_{\hat{\eta}^{\prime}}^{2}\right)
^{2},\label{solB}%
\end{equation}
whereas trace invariance of $M_{H}$ yields%
\begin{equation}
m_{H_{ns}}^{2}+m_{H_{s}}^{2}+2\hat{\mu}_{1/2}^{2}=m_{\eta}^{2}+m_{\hat{\eta}%
}^{2}+m_{\eta^{\prime}}^{2}+m_{\hat{\eta}^{\prime}}^{2}.\label{traceMH}%
\end{equation}
The scalar mass matrix in Eq. (\ref{MS}) can be diagonalized in a similar way.
The $SO(2)\otimes SO(2)\otimes SO(2)$ matrix diagonalizing it is written as%
\begin{equation}
R(\gamma,\delta,\delta^{\prime})=R_{2}(\delta,\delta^{\prime})R_{1}(\gamma)
\end{equation}
with $\gamma,$ $\delta$ and $\delta^{\prime}$ denoting the mixing angles in
the scalar sector analogous to $\alpha,\beta,\beta^{\prime}$ in the
pseudoscalar sector. The former are related to meson masses as%
\begin{align}
\cos2\gamma &  =\frac{m_{S_{ns}}^{2}-m_{S_{s}}^{2}}{m_{\Sigma}^{2}-m_{F}^{2}%
},\qquad\sin2\gamma=\frac{-2m_{S_{s-ns}}^{2}}{m_{\Sigma}^{2}-m_{F}^{2}},\\
\cos2\delta &  =\frac{m_{\Sigma}^{2}-\hat{\mu}^{2}}{m_{\sigma}^{2}%
-m_{\hat{\sigma}}^{2}},\qquad\,\;\sin2\delta=\frac{\epsilon^{2}}%
{m_{\hat{\sigma}}^{2}-m_{\sigma}^{2}},\\
\cos2\delta^{\prime} &  =\frac{m_{F}^{2}-\hat{\mu}^{2}}{m_{f_{0}}^{2}%
-m_{\hat{f}_{0}}^{2}},\qquad\;\sin2\delta^{\prime}=\frac{\epsilon^{2}}%
{m_{\hat{f}_{0}}^{2}-m_{f_{0}}^{2}},
\end{align}
where the physical masses are given by%
\begin{align}
m_{\sigma(\hat{\sigma})}^{2} &  =\frac{1}{2}\left\{  m_{\Sigma}^{2}+\hat{\mu
}_{1/2}^{2}-(+)\sqrt{\left(  m_{\Sigma}^{2}-\hat{\mu}_{1/2}^{2}\right)
^{2}+\epsilon^{4}}\right\}  ,\\
m_{f_{0}(\hat{f}_{0})}^{2} &  =\frac{1}{2}\left\{  m_{F}^{2}+\hat{\mu}%
_{1/2}^{2}-(+)\sqrt{\left(  m_{F}^{2}-\hat{\mu}_{1/2}^{2}\right)
^{2}+\epsilon^{4}}\right\}  ,
\end{align}
with%
\begin{equation}
m_{\Sigma(F)}^{2}=\frac{1}{2}\left\{  m_{S_{ns}}^{2}+m_{S_{s}}^{2}%
-(+)\sqrt{\left(  m_{S_{ns}}^{2}-m_{S_{s}}^{2}\right)  ^{2}+4m_{s_{s-ns}}^{4}%
}\right\}  .
\end{equation}
Summarizing the physical pseudoscalar fields are given by%
\begin{equation}
\left(
\begin{array}
[c]{c}%
\eta\\
\eta^{\prime}\\
\hat{\eta}\\
\hat{\eta}^{\prime}%
\end{array}
\right)  =\left(
\begin{array}
[c]{cccc}%
c_{\alpha}c_{\beta} & -s_{\alpha}c_{\beta} & -c_{\alpha}s_{\beta} & s_{\alpha
}s_{\beta}\\
s_{\alpha}c_{\beta^{\prime}} & c_{\alpha}c_{\beta^{\prime}} & s_{\alpha
}s_{\beta^{\prime}} & c_{\alpha}s_{\beta^{\prime}}\\
c_{\alpha}s_{\beta} & -s_{\alpha}s_{\beta} & c_{\alpha}c_{\beta} & -s_{\alpha
}c_{\beta}\\
-s_{\alpha}s_{\beta}^{\prime} & -c_{\alpha}c_{\beta^{\prime}} & s_{\alpha
}c_{\beta^{\prime}} & c_{\alpha}c_{\beta^{\prime}}%
\end{array}
\right)  \left(
\begin{array}
[c]{c}%
{\small H}_{ns}\\
{\small H}_{s}\\
{\small \hat{H}}_{ns}\\
{\small \hat{H}}_{s}%
\end{array}
\right)  ,\label{explcont}%
\end{equation}
where $c_{\alpha}\equiv\cos\alpha$, $s_{\alpha}\equiv\sin\alpha$. Similar
relations are also valid for the scalar sector with obvious replacements. We
denote as $\sigma,$ $f_{0}$, $\hat{\sigma},$ and  $\hat{f}_{0}$ to the
physical isosinglet scalar fields analogous to $\eta,\ \eta^{\prime}%
,\ \hat{\eta},$ and $\hat{\eta}^{\prime}$.

\subsection{Weak decay constants.}

The vacuum to pseudoscalar matrix elements of axial currents are%
\begin{equation}
\left\langle 0\right\vert A_{a}^{\mu}\left\vert P_{b}(q)\right\rangle
=i\delta_{ab}f_{b}(q^{2})q^{\mu}%
\end{equation}
which in turn imply%
\begin{equation}
\left\langle 0\right\vert \partial_{\mu}A_{a}^{\mu}\left\vert P_{a}%
(q)\right\rangle =f_{a}(q^{2})m_{a}^{2}.\label{fc}%
\end{equation}
We calculate the divergences for the isovector and isospinor axial currents in
our model as%
\begin{equation}
\partial_{\mu}\overrightarrow{A}_{I=1}^{\mu}=2m\left(  b_{0}\vec{P}+\left(
\hat{b}_{0}+m_{s}\hat{d}\right)  \overrightarrow{\hat{P}}\right)
+q.t.,\label{PCACpi}%
\end{equation}%
\begin{equation}
\partial_{\mu}A_{I=1/2}^{\mu}=\left(  m+m_{s}\right)  \left(  b_{0}X+\left(
\hat{b}_{0}+m\hat{d}\right)  \hat{X}\right)  +q.t.,\label{PCACK}%
\end{equation}
where $q.t.$ denote quadratic terms coming from the term $\left\langle
\mathcal{M}_{Q}(\hat{\Phi}\hat{\Phi}^{\dagger}+\hat{\Phi}^{\dagger}\hat{\Phi
})\right\rangle $ in Eq. (\ref{L4q}) and which are irrelevant for the physical
quantities discussed here. Using Eq. (\ref{Cero}) the last equations can be
rewritten to
\[
\partial_{\mu}\overrightarrow{A}_{I=1}^{\mu}=\sqrt{2}a\left(  m_{P}^{2}\vec
{P}-\frac{\epsilon^{2}}{2}\overrightarrow{\hat{P}}\right)  ,\qquad
\partial_{\mu}A_{I=1/2}^{\mu}=\frac{a+b}{\sqrt{2}}\left(  m_{X}^{2}%
X-\frac{\epsilon^{2}}{2}\hat{X}\right)  .
\]
In terms of physical masses and mixing angles we obtain%
\begin{equation}
\partial_{\mu}\overrightarrow{A}_{I=1}^{\mu}=f_{\pi}m_{\pi}^{2}\overrightarrow
{\pi}+f_{\pi}m_{\hat{\pi}}^{2}\tan\theta_{1}\overrightarrow{\hat{\pi}}%
,\qquad\partial_{\mu}A_{I=1/2}^{\mu}=f_{K}m_{K}^{2}K+f_{K}m_{\hat{K}}^{2}%
\tan\theta_{1/2}\hat{K}\label{PCACPI1}%
\end{equation}
where upon using (\ref{fc}) we identified%
\begin{equation}
a=\frac{f_{\pi}}{\sqrt{2}\cos\theta_{1}},\qquad a+b=\frac{\sqrt{2}f_{K}}%
{\cos\theta_{1/2}}.\label{amb1}%
\end{equation}

\section{Predictions: meson masses.}

There are eight free parameters in the model: $\{\mu^{2},$ $\lambda,$ $B,$
$\epsilon^{2},$ $a,$ $b,$ $\hat{\mu}_{1}^{2},$ $\hat{\mu}_{1/2}^{2}\}$%
\emph{.}\ The parameters $\hat{\mu}_{1/2}^{2}$ and $\epsilon^{4}$ are given in
terms of physical masses in Eqs. (\ref{mug}, \ref{eps}). Eqs. (\ref{solB},
\ref{traceMH}, \ref{ma}, \ref{m12}, \ref{amb1}), allow us to fix the remaining
parameters. As input we use the physical masses for $\pi(137)$, $a_{0}(980)$,
$K(495),$ $\eta(548),$ $\eta^{\prime}(957),$ $\eta(1295)$ in addition to the
weak decay constants $f_{\pi}$ and $f_{K}$. The vev's $a$, $b$ can be written
in terms of these physical quantities as follows. From Eqs. (\ref{mpi},
\ref{ma}, \ref{m12}) we get%
\begin{align}
m_{P}^{2} &  =m_{\pi}^{2}+\frac{\epsilon^{4}}{4(\hat{\mu}_{1}^{2}-m_{\pi}%
^{2})},\qquad\quad m_{\hat{\pi}}^{2}=\hat{\mu}_{1}^{2}+\frac{\epsilon^{4}%
}{4(\hat{\mu}_{1}^{2}-m_{\pi}^{2})},\label{mpmpi}\\
m_{X}^{2} &  =m_{K}^{2}+\frac{\epsilon^{4}}{4(\hat{\mu}_{1/2}^{2}-m_{K}^{2}%
)},\qquad m_{\hat{K}}^{2}=\hat{\mu}_{1/2}^{2}+\frac{\epsilon^{4}}{4(\hat{\mu
}_{1/2}^{2}-m_{K}^{2})},\label{mxmk}%
\end{align}
and from these and Eqs. (\ref{theta1}, \ref{theta2})%
\begin{align}
\left(  \cos\theta_{1}\right)  ^{-1} &  =\frac{1}{2}\left[  4+\frac
{\epsilon^{4}}{\left(  m_{\pi}^{2}-\hat{\mu}_{1}^{2}\right)  ^{2}}\right]
^{1/2},\\
\left(  \cos\theta_{1/2}\right)  ^{-1} &  =\frac{1}{2}\left[  4+\frac
{\epsilon^{4}}{\left(  m_{K}^{2}-\hat{\mu}_{1/2}^{2}\right)  ^{2}}\right]
^{1/2},
\end{align}
which are functions of the physical masses, $\hat{\mu}_{1}^{2}$ and $\hat{\mu
}_{1/2}^{2}$ through Eqs. (\ref{mug}, \ref{eps}). In a similar way, for the
scalar sector and $m_{\hat{\eta}^{\prime}}^{2}$ we obtain%
\begin{equation}
m_{S}^{2}=m_{a}^{2}+\frac{\epsilon^{4}}{4(\hat{\mu}_{1}^{2}-m_{a}^{2})},\qquad
m_{A}^{2}=\hat{\mu}_{1}^{2}+\frac{\epsilon^{4}}{4(\hat{\mu}_{1}^{2}-m_{a}%
^{2})},\qquad m_{\hat{\eta}^{\prime}}^{2}=\hat{\mu}_{1/2}^{2}+\frac
{\epsilon^{4}}{4(\hat{\mu}_{1/2}^{2}-m_{\eta^{\prime}}^{2})}.
\end{equation}
These relations and Eqs. (\ref{amb1}) yield $a$ and $b$ as functions of the
physical masses, $\hat{\mu}_{1}^{2}$ and $\hat{\mu}_{1/2}^{2}$

Some useful relations obtained from Eqs. (\ref{mp}, \ref{mHNS}, \ref{mHS},
\ref{traceMH}) are%
\begin{equation}
\lambda(a^{2}-b^{2})=-6Bb-(m_{\eta}^{2}+m_{\eta^{\prime}}^{2}+m_{\hat{\eta}%
}^{2}+m_{\hat{\eta}^{\prime}}^{2})+2\hat{\mu}_{1/2}^{2}+2m_{p}^{2},
\end{equation}%
\begin{equation}
m_{H_{ns}}^{2}-m_{H_{s}}^{2}=-2Bb+\lambda(a^{2}-b^{2}),
\end{equation}%
\begin{equation}
\xi=m_{p}^{2}-2Bb-\lambda a^{2}%
\end{equation}
or%
\begin{equation}
m_{H_{ns}}^{2}-m_{H_{s}}^{2}=-8Bb-(m_{\eta}^{2}+m_{\eta^{\prime}}^{2}%
+m_{\hat{\eta}}^{2}+m_{\hat{\eta}^{\prime}}^{2})+2\hat{\mu}_{1/2}^{2}%
+2m_{p}^{2}.
\end{equation}
Using the latter equation and Eq. (\ref{solB}) we obtain%
\begin{gather}
8B^{2}(a^{2}+2b^{2})-4Bb\left\{  2(\hat{\mu}_{1/2}^{2}+m_{p}^{2})-(m_{\eta
}^{2}+m_{\eta^{\prime}}^{2}+m_{\hat{\eta}}^{2}+m_{\hat{\eta}^{\prime}}%
^{2})\right\}  +\nonumber\\
\left\{  \hat{\mu}_{1/2}^{2}+m_{p}^{2}-(m_{\eta}^{2}+m_{\hat{\eta}}%
^{2})\right\}  \left\{  \hat{\mu}_{1/2}^{2}+m_{p}^{2}-(m_{\eta^{\prime}}%
^{2}+m_{\hat{\eta}^{\prime}}^{2})\right\}  =0,
\end{gather}
with the solution%
\begin{equation}
B=\frac{Y\pm\sqrt{Y^{2}-32Z\left(  a^{2}+2b^{2}\right)  }}{16\left(
a^{2}+2b^{2}\right)  }\label{B}%
\end{equation}
where%
\begin{align}
Y &  =4b\left\{  2(\hat{\mu}_{1/2}^{2}+m_{p}^{2})-(m_{\eta}^{2}+m_{\eta
^{\prime}}^{2}+m_{\hat{\eta}}^{2}+m_{\hat{\eta}^{\prime}}^{2})\right\}  ,\\
Z &  =\left\{  \hat{\mu}_{1/2}^{2}+m_{p}^{2}-(m_{\eta}^{2}+m_{\hat{\eta}}%
^{2})\right\}  \left\{  \hat{\mu}_{1/2}^{2}+m_{p}^{2}-(m_{\eta^{\prime}}%
^{2}+m_{\hat{\eta}^{\prime}}^{2})\right\}  .
\end{align}

These relations yield the parameters in terms of physical quantities and
$\hat{\mu}_{1}^{2}$, $\hat{\mu}_{1/2}^{2}$. A numerical solution for $\hat
{\mu}_{1}^{2}$ and $\hat{\mu}_{1/2}^{2}$ can be found from the following
equations
\begin{align}
m_{S}^{2}-m_{p}^{2} &  =-4Bb+2\lambda a^{2},\\
m_{X}^{2}-m_{p}^{2} &  =\left(  -2B+\lambda\right)  (b-a).
\end{align}
These equations depend only on $\hat{\mu}_{1}^{2}$ and $\hat{\mu}_{1/2}^{2}$
via Eqs. (\ref{amb1}-\ref{B}) and Eq. (\ref{eps}).

The input used, and the values for the parameters, are listed in Table I. The
numerical solution yield $\hat{\mu}_{1}=1257.5$ MeV and $\hat{\mu}%
_{1/2}=1206.2$ MeV for the central values in the input. These quantities are
the central values for the masses of the pure isotriplet and isodoublet
four-quark states in the presence of the appropriate quark mass terms
$\mathcal{M}_{Q}$ in Eq. (\ref{L4q}) and are valid to leading order in the
SU(3) breaking $\Delta$ in Eq. (\ref{Delta}). The results are consistent with
the expected scale and also with an inverted spectrum for tetraquarks.

\begin{center}
Table I

$%
\begin{tabular}
[c]{||l|l|l|l||}\hline\hline
& Input & Parameter & Fit\\\hline
$m_{\pi}$ & $137.3\pm2.3\ $MeV & $\hat{\mu}_{1}(${\small MeV}$)$ &
$1257.5\pm61.3$\\\hline
$m_{a}$ & $984.3\pm0.9\ $MeV & $\hat{\mu}_{1/2}(${\small MeV}$)$ &
$1206.2\pm142.3$\\\hline
$f_{\pi}$ & $92.42\pm3.53\ $MeV & $\left\vert \epsilon\right\vert
(${\small MeV}$)$ & $1012.3\pm75.9$\\\hline
$m_{K}$ & $495.67\pm2.00\ $MeV & $a(${\small MeV}$)$ & $68.78\pm4.47$\\\hline
$f_{K}$ & $113.0\pm1.3\ $MeV & $b(${\small MeV}$)$ & $104.8\pm2.6$\\\hline
$m_{\eta}$ & $547.3\pm0.12\ $MeV & $B(${\small GeV}$)$ & $-2.16\pm
0.28$\\\hline
$m_{\eta^{\prime}}$ & $957.78\pm0.14\ $MeV & $\lambda$ & $31.8\pm5.9$\\\hline
$m_{\hat{\eta}}$ & $1297\pm2.8\ $MeV & $\mu^{2}(${\small GeV}$^{2})$ &
$0.490\pm0.107$\\\hline\hline
\end{tabular}
$
\end{center}

Uncertainties listed in Table I correspond to the measured values in the case
of the isosinglets \cite{PDG}. Since we are working in the good isospin limit
we use the experimental deviations from this limit in the case of isovectors
and isodoublets. As for $f_{K}$, we have no information for the neutral case
hence we take the value for this constant from the charged case and its
uncertainties. Using these values for the input we fix the parameters of the
model to the values listed in Table I.\ Predictions for the remaining meson
masses and mixing angles are shown in Table II. Results are consistent with
the identification of the physical mesons with the $\eta(1295),$ $\eta(1450),$
$K(1469),$ $\pi(1300),$ as well as the following scalar states: $f_{0}(1370),$
$f_{0}(1500),$ $K_{0}^{\ast}(1430),$ $a_{0}(1450),$ $f_{0}(600),$
$f_{0}(980),$ $f_{0}(1500)$.

\begin{center}
Table II: Predictions

$%
\begin{tabular}
[c]{||l|l|l|l|l|l||}\hline\hline
Mass & Prediction & Identification & Experiment & Angle & Prediction\\\hline
$m_{\hat{\pi}}($MeV$)$ & $1322.6\pm32.4$ & $\pi(1300)$ & $1300\pm
200$\cite{PDG} & $\theta_{1}$ & $18.16^{\circ}\pm4.3^{\circ}$\\\hline
$m_{\hat{K}}($MeV$)$ & $1293.1\pm3.5$ & $K(1469)$ & $1400-1460$\cite{PDG} &
$\theta_{1/2}$ & $22.96^{\circ}\pm4.8^{\circ}$\\\hline
$m_{A}($MeV$)$ & $1417.3\pm51.0$ & $a_{0}(1450)$ & $1474\pm19$\cite{PDG} &
$\phi_{1}$ & $39.8^{\circ}\pm4.5^{\circ}$\\\hline
$m_{\kappa}($MeV$)$ & $986.2\pm19.1$ & $\kappa(900)$ & $750-950$%
\cite{E791kappa,Svec,Syracusekappa} & $\phi_{1/2}$ & $46.7^{\circ}%
\pm9.5^{\circ}$\\\hline
$m_{\hat{\kappa}}($MeV$)$ & $1413.9\pm76.4$ & $K_{0}^{\ast}(1430)$ &
$1429\pm4\pm5$\cite{PDG} & $\alpha$ & $53.4^{\circ}\pm0.8^{\circ}$\\\hline
$m_{\hat{\eta}}^{\prime}($MeV$)$ & $1394.0\pm61.9$ & $\eta(1450)$ &
$1400-1470$\cite{PDG} & $\beta$ & $23.9^{\circ}\pm5.1^{\circ}$\\\hline
$m_{\sigma}($MeV$)$ & $380.6\pm91.0$ & $f_{0}(600)$ & $478$ $\pm$
$35$\cite{PDG} & $\beta^{\prime}$ & $43.6^{\circ}\pm7.2^{\circ}$\\\hline
$m_{f_{0}}($MeV$)$ & $1022.4\pm25.6$ & $f_{0}(980)$ & $980\pm10$\cite{PDG} &
$\gamma$ & $-9.11^{\circ}\pm0.5^{\circ}$\\\hline
$m_{\hat{\sigma}}($MeV$)$ & $1284.7\pm15.3$ & $f_{0}(1370)$ & $1200-1500$%
\cite{PDG} & $\delta$ & $21.45^{\circ}\pm6.5^{\circ}$\\\hline
$m_{\hat{f}_{0}}($MeV$)$ & $1447.7\pm84.6$ & $f_{0}(1500)$ & $1500\pm
10$\cite{PDG} & $\delta^{\prime}$ & $51.36^{\circ}\pm8.3^{\circ}%
$\\\hline\hline
\end{tabular}
$
\end{center}

\begin{figure}
[h]
\begin{center}
\includegraphics[
height=2.4647in,
width=3.9055in
]%
{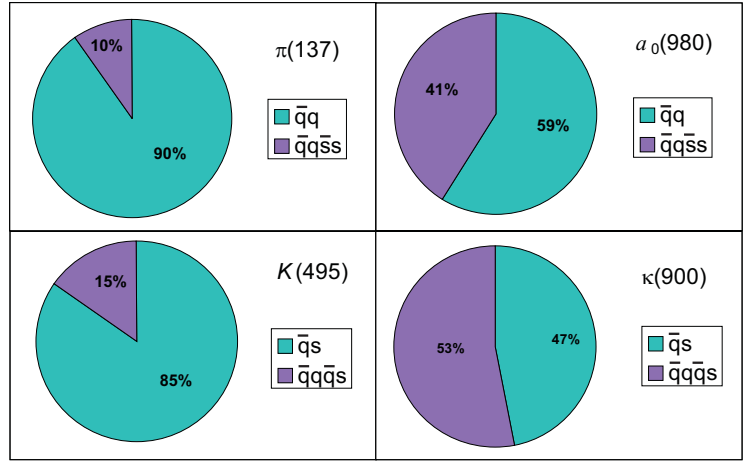}%
\caption{quark content of light isotriplets and isodoublets.}%
\label{qcont}%
\end{center}
\end{figure}

\begin{figure}
[hptb]
\begin{center}
\includegraphics[
height=2.4751in,
width=3.9202in
]%
{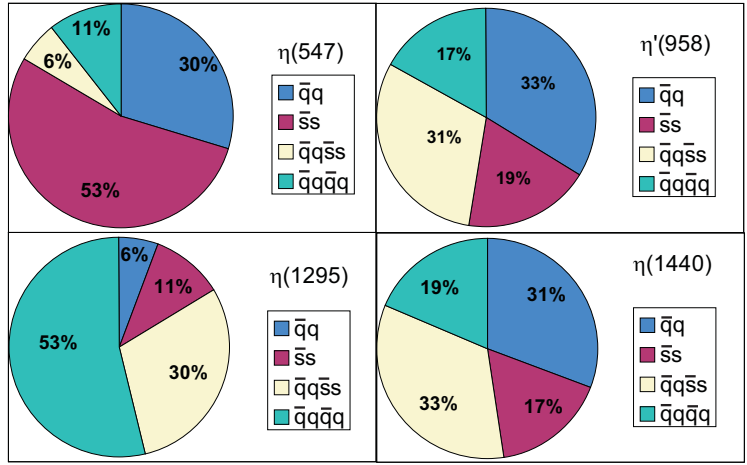}%
\caption{quark content of pseudoscalar isosinglets }%
\label{etas}%
\end{center}
\end{figure}

\begin{figure}
[ptb]
\begin{center}
\includegraphics[
height=2.5192in,
width=3.9989in
]%
{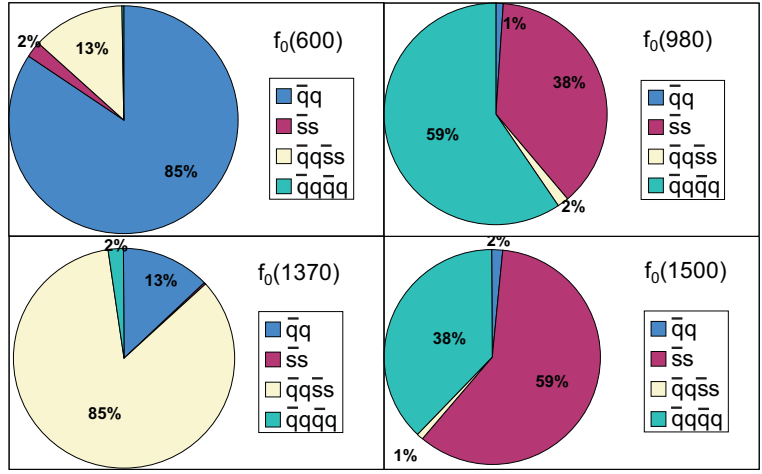}%
\caption{quark content of scalar isosinglets }%
\label{f0s}%
\end{center}
\end{figure}

The quark content of mesons corresponding to the central values of the mixing
angles are shown in Figs. \ref{qcont}, \ref{etas}, \ref{f0s}. Light
isodoublets and isotriples are shown in Fig. \ref{qcont}, heavy mesons being
composites with the opposite quark content.%
Figure \ref{etas} show the quark contents of the isoscalar-pseudoscalar mesons
whereas the quark contents of isoscalar-scalar mesons is shown in figure
\ref{f0s}.%
%

It is worth to remark that the picture for isosinglets can be modified by the
introduction of glueball fields which are expected to mix with isosinglet
four-quark and $\overline{q}q$\ states.

The light isotriplet and isodoublets, pions and kaons, arise as states mainly
composed of $q\overline{q}$, whereas the heavy fields $\pi(1300),$ $K(1460)$
arise as mainly tetraquark states. In contrast, in the scalar sector
isotriplets and isodoublets get strongly mixed and the physical mesons turn
out to have almost identical amounts of $q\overline{q}$ and four-quark
content. In the case of $\eta(547)$ we obtain also a small four-quark content,
the $\eta(1295)$ being almost a four-quark state. However, the $\eta^{\prime
}(958)$ and $\eta(1440),$ turn out to be strong admixtures of $q\overline{q}$
and four-quark states with almost equal amounts of them. As to the isosinglet
scalars, the sigma meson ( $f_{0}(600)$) arises as mainly $q\overline{q}$
state and the opposite quark content is carried by the $f_{0}(1370)$ which
turns out to be mainly a four quark state. Finally, the $f_{0}(980)$ turns out
to have roughly a 40\% content of $\overline{q}q$ (more explicitly
$\overline{s}s$ ) and 60\% of $\overline{q}\overline{q}qq$ whereas the
$f_{0}(1500)$ is composed 60\% of $\overline{q}q$ (more explicitly
$\overline{s}s$ ) and almost 40\% of $\overline{q}\overline{q}qq$. As pointed
above, the quark content of isoscalar mesons as arising in this model are
expected to be modified by the inclusion of glueballs in the present context,
thus the picture for isoscalar mesons as arising from this analysis should be
taken as very preliminary results. The need for the inclusion of glueball
fields will be more transparent when we analyze interactions in the next section.

\section{Interactions.}

As shown in the previous section, the meson spectrum is consistent with the
mechanisms conjectured in this work. Another important piece of information on
meson structure is the coupling to the different allowed channels. We work out
the predictions of the model for the most relevant three-meson couplings and
compare the predicted decay widths with experimental results for those decays
where confident results exist. The different pieces in Eq. (\ref{L3}) relevant
to the couplings we are interested are shown in Table III in terms of the
parameters of the model.

\begin{center}
Table III%

\[%
\begin{tabular}
[c]{||l|l|l||}\hline\hline
Lagrangian & Explicit form & Couplings involved\\\hline
$\mathcal{L}_{S^{0}X^{+}X^{-}}$ & $\frac{1}{\sqrt{2}}\left(  2B-\left(
2a-b\right)  \lambda\right)  S^{0}X^{+}X^{-}$ & $g_{a^{0}K^{+}K^{-}},$
$g_{AK^{+}K^{-}}$\\\hline
$\mathcal{L}_{S^{0}Y^{+}Y^{-}}$ & $-\frac{1}{\sqrt{2}}\left(  2B+\left(
2a+b\right)  \lambda\right)  S^{0}Y^{+}Y^{-}$ & $g_{a^{0}\kappa^{+}\kappa^{-}%
},$ $g_{A\kappa^{+}\kappa^{-}}$\\\hline
$\mathcal{L}_{\tilde{F}\pi^{-}\pi^{+}}$ & $-\left(  \sqrt{2}\lambda
aS_{ns}+2BS_{s}\right)  P^{+}P^{-}$ & $g_{\sigma\pi^{+}\pi^{-}},$ $g_{f\pi
^{+}\pi^{-}}$\\\hline
$\mathcal{L}_{FX^{+}X^{-}}$ & $-\frac{1}{\sqrt{2}}\left[  \lambda\left(
2a-b\right)  \lambda+2B\right]  S_{ns}X^{+}X^{-}-\left[  \lambda\left(
2b-a\right)  \right]  S_{s}X^{+}X^{-}$ & $g_{\sigma KK},$ $g_{fKK}$\\\hline
$\mathcal{L}_{FY^{+}Y^{-}}$ & $-\frac{1}{\sqrt{2}}\left[  \lambda\left(
2a+b\right)  \lambda-2B\right]  S_{ns}Y^{+}Y^{-}-\left[  \lambda\left(
2b+a\right)  \right]  S_{s}Y^{+}Y^{-}$ & $g_{\sigma\kappa^{+}\kappa^{-}},$
$g_{f\kappa^{+}\kappa^{-}}$\\\hline
$\mathcal{L}_{S^{0}P^{0}H}$ & $-\left(  \sqrt{2}\lambda aH_{ns}+2BH_{s}%
\right)  S^{0}P^{0}$ & $g_{A\pi\eta}$, $g_{A\pi\eta^{\prime}}$\\\hline
$\mathcal{L}_{Y^{0}P^{+}X^{-}}$ & $\left(  -\lambda b+2B\right)  Y^{0}%
P^{+}X^{-}$ & $g_{\hat{\kappa}^{0}\pi^{+}K^{-}}$\\\hline\hline
\end{tabular}
\]

\end{center}

In order to exhibit the manipulations involved let us explicitly calculate the
$g_{\sigma\pi\pi}$ coupling. The relevant piece $\mathcal{L}_{\tilde{F}\pi\pi
}$ in Table III can be rewritten to%
\begin{equation}
\mathcal{L}_{\tilde{F}\pi\pi}=-\frac{1}{\sqrt{2}a}\left[  (m_{S_{ns}}%
^{2}-m_{P}^{2})S_{ns}+\frac{1}{2}\left(  m_{F}^{2}-m_{\Sigma}^{2}\right)
\sin2\gamma S_{s}\right]  P^{+}P^{-}.
\end{equation}
It is possible to rewrite this Lagrangian in terms of the $\Sigma$ and $F$
fields defined as
\begin{align}
S_{ns} &  =\cos\gamma\Sigma+\sin\gamma F,\\
S_{s} &  =-\sin\gamma\Sigma+\cos\gamma F.
\end{align}
We obtain%
\begin{equation}
\mathcal{L}_{\tilde{F}\pi\pi}=-\frac{1}{\sqrt{2}a}\left[  (m_{\Sigma}%
^{2}-m_{P}^{2})\cos\gamma\Sigma+(m_{F}^{2}-m_{P}^{2})\sin\gamma F\right]
P^{+}P^{-}.
\end{equation}
From where , upon the use of the following relations%
\begin{align}
m_{F}^{2} &  =m_{f}^{2}\cos^{2}\delta^{\prime}+m_{\hat{f}}^{2}\sin^{2}%
\delta^{\prime},\\
m_{\Sigma}^{2} &  =m_{\sigma}^{2}\cos^{2}\delta+m_{\hat{\sigma}}^{2}\sin
^{2}\delta,\\
m_{p}^{2} &  =m_{\pi}^{2}\cos^{2}\theta_{1}+m_{\hat{\pi}}^{2}\sin^{2}%
\theta_{1},
\end{align}
and the expressions for $\Sigma$, $F$ and $P$ in terms of the physical fields
we get
\begin{equation}
g_{\sigma\pi\pi}=\frac{1}{\sqrt{2}a}(m_{\sigma}^{2}+m_{\hat{\sigma}}%
^{2}-m_{\pi}^{2}-m_{\hat{\pi}}^{2}+\hat{\mu}_{1}^{2}-\hat{\mu}_{1/2}^{2}%
)\cos\gamma\cos\delta\cos^{2}\theta_{1}.
\end{equation}
Under similar manipulations for the appropriate pieces of the Lagrangian in
Table III we obtain the couplings listed in Table IV.

\begin{center}
Table IV
\end{center}%

\[%
\begin{tabular}
[c]{||l|l|l||}\hline\hline
Coupling & Analytic form & Prediction (GeV)\\\hline
$g_{a_{0}K^{+}K^{-}}$ & $-\frac{1}{\sqrt{2}\left(  a+b\right)  }\left(
m_{K}^{2}+m_{\hat{K}}^{2}-m_{a}^{2}-m_{A}^{2}+\hat{\mu}_{1}^{2}-\hat{\mu
}_{1/2}^{2}\right)  \cos\phi_{1}\cos^{2}\theta_{1/2}$ & $\ \ 2.50\pm
0.57$\\\hline
$g_{a_{0}\kappa^{+}\kappa^{-}}$ & $\frac{1}{\sqrt{2}\left(  b-a\right)
}\left(  m_{\kappa}^{2}+m_{\hat{\kappa}}^{2}-m_{a}^{2}-m_{A}^{2}+\hat{\mu}%
_{1}^{2}-\hat{\mu}_{1/2}^{2}\right)  \cos\phi_{1}\cos^{2}\phi_{1/2}$ &
$-0.44\pm1.45$\\\hline
$g_{a_{0}\pi\eta}$ & $\frac{1}{\sqrt{2}a}\left(  m_{A}^{2}+m_{a}%
^{2}-m_{\widehat{\eta}}^{2}-m_{\eta}^{2}-\hat{\mu}_{1}^{2}+\hat{\mu}_{1/2}%
^{2}\right)  \cos\phi_{1}\cos\alpha\cos\beta\cos\theta_{1}$ & $\ \ 3.83\pm
0.60$\\\hline
$g_{\sigma\pi^{+}\pi^{-}}$ & $\frac{1}{\sqrt{2}a}(m_{\sigma}^{2}%
+m_{\hat{\sigma}}^{2}-m_{\pi}^{2}-m_{\hat{\pi}}^{2}+\hat{\mu}_{1}^{2}-\hat
{\mu}_{1/2}^{2})\cos\gamma\cos\delta\cos^{2}\theta_{1},$ & $\ \ 2.22\pm1.37
$\\\hline
$g_{\sigma K^{+}K^{-}}$ & $\frac{1}{\sqrt{2}(a+b)}\left(  m_{\sigma}%
^{2}+m_{\hat{\sigma}}^{2}-m_{K}^{2}-m_{\hat{K}}^{2}\right)  \left(  \cos
\gamma-\sqrt{2}\sin\gamma\right)  \cos\delta\cos^{2}\theta_{1/2}$ &
$-0.31\pm0.68$\\\hline
$g_{\sigma\kappa^{+}\kappa^{-}}$ & $-\frac{1}{\sqrt{2}(b-a)}\left(  m_{\sigma
}^{2}+m_{\hat{\sigma}}^{2}-m_{\kappa}^{2}-m_{\hat{\kappa}}^{2}\right)  \left(
\cos\gamma+\sqrt{2}\sin\gamma\right)  \cos\delta\cos^{2}\phi_{1/2}$ &
$\ \ 1.42\pm0.89$\\\hline
$g_{f(980)\pi^{+}\pi^{-}}$ & $\frac{1}{\sqrt{2}a}(m_{f}^{2}+m_{\hat{f}}%
^{2}-m_{\pi}^{2}-m_{\hat{\pi}}^{2}+\hat{\mu}_{1}^{2}-\hat{\mu}_{1/2}^{2}%
)\sin\gamma\cos\delta^{\prime}\cos^{2}\theta_{1}$ & $-1.43\pm0.34$\\\hline
$g_{f(980)K^{+}K^{-}}$ & $\frac{1}{\sqrt{2}(a+b)}\left(  m_{f}^{2}+m_{\hat{f}%
}^{2}-m_{K}^{2}-m_{\hat{K}}^{2}\right)  \left(  \sin\gamma+\sqrt{2}\cos
\gamma\right)  \cos\delta^{\prime}\cos^{2}\theta_{1/2}$ & $\ \ 3.27\pm
0.99$\\\hline
$g_{f(980)\kappa^{+}\kappa^{-}}$ & $-\frac{1}{\sqrt{2}(b-a)}\left(  m_{f}%
^{2}+m_{\hat{f}}^{2}-m_{\kappa}^{2}-m_{\hat{\kappa}}^{2}\right)  \left(
\sin\gamma-\sqrt{2}\cos\gamma\right)  \cos\delta^{\prime}\cos^{2}\phi_{1/2}$ &
$\ \ 0.77\pm1.67$\\\hline
$g_{f(1370)\pi^{+}\pi^{-}}$ & $\frac{1}{\sqrt{2}a}(m_{\sigma}^{2}%
+m_{\hat{\sigma}}^{2}-m_{\pi}^{2}-m_{\hat{\pi}}^{2}+\hat{\mu}_{1}^{2}-\hat
{\mu}_{1/2}^{2})\cos\gamma\sin\delta\cos^{2}\theta_{1}$ & $\ \ 0.87\pm
0.77$\\\hline
$g_{f(1370)K^{+}K^{-}}$ & $\frac{1}{\sqrt{2}(a+b)}\left(  m_{\sigma}%
^{2}+m_{\hat{\sigma}}^{2}-m_{K}^{2}-m_{\hat{K}}^{2}\right)  \left(  \cos
\gamma-\sqrt{2}\sin\gamma\right)  \sin\delta\cos^{2}\theta_{1/2}$ &
$-0.12\pm0.24$\\\hline
$g_{f(1370)\kappa^{+}\kappa^{-}}$ & $-\frac{1}{\sqrt{2}(b-a)}\left(
m_{\sigma}^{2}+m_{\hat{\sigma}}^{2}-m_{\kappa}^{2}-m_{\hat{\kappa}}%
^{2}\right)  \left(  \cos\gamma+\sqrt{2}\sin\gamma\right)  \sin\delta\cos
^{2}\phi_{1/2}$ & $\ \ 0.56\pm0.32$\\\hline
$g_{f(1500)\pi^{+}\pi^{-}}$ & $\frac{1}{\sqrt{2}a}(m_{f}^{2}+m_{\hat{f}}%
^{2}-m_{\pi}^{2}-m_{\hat{\pi}}^{2}+\hat{\mu}_{1}^{2}-\hat{\mu}_{1/2}^{2}%
)\sin\gamma\sin\delta^{\prime}\cos^{2}\theta_{1}$ & $-1.80\pm0.58$\\\hline
$g_{f(1500)K^{+}K^{-}}$ & $\frac{1}{\sqrt{2}(a+b)}\left(  m_{f}^{2}+m_{\hat
{f}}^{2}-m_{K}^{2}-m_{\hat{K}}^{2}\right)  \left(  \sin\gamma+\sqrt{2}%
\cos\gamma\right)  \sin\delta^{\prime}\cos^{2}\theta_{1/2}.$ & $\ \ 4.09\pm
1.64$\\\hline
$g_{f(1500)\kappa^{+}\kappa^{-}}$ & $-\frac{1}{\sqrt{2}(b-a)}\left(  m_{f}%
^{2}+m_{\hat{f}}^{2}-m_{\kappa}^{2}-m_{\hat{\kappa}}^{2}\right)  \left(
\sin\gamma-\sqrt{2}\cos\gamma\right)  \sin\delta^{\prime}\cos^{2}\phi_{1/2}.$
& $\ \ 0.96\pm3.75$\\\hline
$g_{A_{0}K^{+}K^{-}}$ & $-\frac{1}{\sqrt{2}\left(  a+b\right)  }\left(
m_{K}^{2}+m_{\hat{K}}^{2}-m_{a}^{2}-m_{A}^{2}+\hat{\mu}_{1}^{2}-\hat{\mu
}_{1/2}^{2}\right)  \sin\phi_{1}\cos^{2}\theta_{1/2}$ & $\ \ 2.08\pm
0.62$\\\hline
$g_{A_{0}\kappa^{+}\kappa^{-}}$ & $\frac{1}{\sqrt{2}\left(  b-a\right)
}\left(  m_{\kappa}^{2}+m_{\hat{\kappa}}^{2}-m_{a}^{2}-m_{A}^{2}+\hat{\mu}%
_{1}^{2}-\hat{\mu}_{1/2}^{2}\right)  \sin\phi_{1}\cos^{2}\phi_{1/2}$ &
$-0.36\pm1.23$\\\hline
$g_{A\pi\eta\text{ }}$ & $\frac{1}{\sqrt{2}a}\left(  m_{A}^{2}+m_{a}%
^{2}-m_{\widehat{\eta}}^{2}-m_{\eta}^{2}-\hat{\mu}_{1}^{2}+\hat{\mu}_{1/2}%
^{2}\right)  \sin\phi_{1}\cos\alpha\cos\beta\cos\theta_{1}$ & $\ \ 3.19\pm
0.79$\\\hline
$g_{A\pi\eta^{\prime}\text{ }}$ & $\frac{1}{\sqrt{2}a}\left(  m_{A}^{2}%
+m_{a}^{2}-m_{\widehat{\eta}^{\prime}}^{2}-m_{\eta^{\prime}}^{2}-\hat{\mu}%
_{1}^{2}+\hat{\mu}_{1/2}^{2}\right)  \sin\phi_{1}\sin\alpha\cos\beta^{\prime
}\cos\theta_{1}$ & $\ \ 0.0046\pm0.4837$\\\hline
$g_{\hat{\kappa}^{0}\pi^{+}K^{-}}$ & $\frac{1}{2b}\left\{  m_{\hat{K}}%
^{2}+m_{K}^{2}+m_{\hat{\kappa}}^{2}+m_{\kappa}^{2}-2(m_{\hat{\pi}}^{2}+m_{\pi
}^{2}-\hat{\mu}_{1}^{2}+\hat{\mu}_{1/2}^{2})\right\}  \sin\phi_{1/2}\cos
\theta_{1}\cos\theta_{1/2}$ & $\ \ 4.68\pm1.55$\\\hline\hline
\end{tabular}
\]
Notice that in the absence of quark mass terms for tetraquarks ( $\hat{\mu
}_{1}^{2}=\hat{\mu}_{1/2}^{2}$ ) and and the $\epsilon^{2}$ term, all the
fields in the \textquotedblleft nonstandard\textquotedblright\ nonet have a
common mass $\hat{\mu}^{2}$ hence the couplings $g_{a_{0}K^{+}K^{-}}$,
$g_{a_{0}\kappa^{+}\kappa^{-}}$ reduce to those obtained in the linear sigma
model \cite{Mauro,TornqvistEJP}%
\begin{align}
g_{a_{0}K^{+}K^{-}}^{L\sigma M} &  =\frac{1}{\sqrt{2}\left(  a+b\right)
}\left(  m_{a}^{2}-m_{K}^{2}\right)  ,\\
g_{a_{0}\kappa^{+}\kappa^{-}}^{L\sigma M} &  =-\frac{1}{\sqrt{2}\left(
b-a\right)  }\left(  m_{a}^{2}-m_{\kappa}^{2}\right)  .
\end{align}
The modifications to these couplings come from the mixing of the fields and
from quark mass terms.

Unfortunately there is no much experimental information on these couplings.
Recently, KLOE Collaboration \cite{kloe1} \cite{kloe2} at DA$\Phi$NE measured
the di-meson mass spectrum in the rare $\varphi\longrightarrow\pi^{0}\pi
^{0}\gamma$ and $\varphi\longrightarrow\pi\eta\gamma$ decays. The couplings
$g_{a_{0}K^{+}K^{-}}$, $g_{f_{0}K^{+}K^{-}}$ and $g_{f_{0}\pi^{+}\pi^{-}}$ as
extracted from these experiments are
\begin{align*}
\frac{g_{f_{0}K^{+}K^{-}}^{2}}{4\pi}|_{KLOE}  & =2.79\pm0.12\text{ GeV}%
^{2},\qquad\frac{g_{f_{0}K^{+}K^{-}}^{2}}{g_{f_{0}\pi^{+}\pi^{-}}^{2}}%
|_{KLOE}=4.00\pm0.14,\\
\frac{g_{a_{0}K^{+}K^{-}}^{2}}{4\pi}|_{KLOE}  & =0.40\pm0.04\text{ GeV}%
^{2},\qquad\frac{g_{a_{0}\pi^{0}\eta}^{2}}{g_{a_{0}K^{+}K^{-}}^{2}}%
|_{KLOE}=1.35\pm0.09,\qquad\frac{g_{f_{0}KK}^{2}}{g_{a_{0}KK}^{2}}%
|_{KLOE}=7.0\pm0.7.
\end{align*}
The predictions of the model for these quantities are%
\begin{align*}
\frac{g_{f_{0}K^{+}K^{-}}^{2}}{4\pi}|_{\text{Model}}  & =0.85\pm0.52\text{
GeV}^{2},\qquad\frac{g_{f_{0}K^{+}K^{-}}^{2}}{g_{f_{0}\pi^{+}\pi^{-}}^{2}%
}|_{\text{Model}}=5.16\pm3.96,\\
\frac{g_{a_{0}K^{+}K^{-}}^{2}}{4\pi}|_{\text{Model}}  & =0.50\pm0.23\text{
GeV}^{2},\qquad\frac{g_{a_{0}\pi^{0}\eta}^{2}}{g_{a_{0}K^{+}K^{-}}^{2}%
}|_{\text{Model}}=2.34\pm1.30,\qquad\frac{g_{f_{0}KK}^{2}}{g_{a_{0}KK}^{2}%
}|_{\text{Model}}=1.71\pm1.30.
\end{align*}

These results give valuable information. On the one side, our $g_{a_{0}%
K^{+}K^{-}\text{ }}$ is in good agreement with data and the ratio
$\frac{g_{a_{0}\pi^{0}\eta}^{2}}{g_{a_{0}K^{+}K^{-}}^{2}}$ is consistent with
experimental results. The first quantity involves only the isodoublet and
isotriplet sectors. The second one involve only the $\eta$ meson from the
isoscalar sector. The prediction of the model for the isodoublet and
isotriplet sectors will hardly be substantially modified by the introduction
of glueball fields which are expected to impact mainly in the isoscalar
sectors. In the latter sector we expect only slight modifications to the
predictions for the $\eta$ meson due to its relatively low mass. On the other
side, the predictions of the model for $g_{f_{0}K^{+}K^{-}}$ is smaller than
KLOE measurement roughly by a factor $\sqrt{3}$. In spite of this, the ratio
$\frac{g_{f_{0}K^{+}K^{-}}^{2}}{g_{f_{0}\pi^{+}\pi^{-}}^{2}}$ is in good
agreement with data. A look on the analytic form of the couplings in Table IV
reveals that both $g_{f_{0}K^{+}K^{-}}$ and $g_{f_{0}\pi^{+}\pi^{-}}$ depend
on $\delta^{\prime}$ as $cos\delta^{\prime}.$ This factor cancels in the
ratio. Furthermore this mixing angle turns out to be \ unexpectedly large in
the fit . We expect that upon the introduction of the lightest scalar glueball
in the model this mixing angle get modified yielding more reasonable values
for the couplings of the $f_{0}(980)$ meson. According to the scalar analogous
to Eq. (\ref{explcont}) the lowering of this mixing angle will also reduce the
four-quark content in this meson. In this sense, the picture of isoscalar
mesons as arising in this work must be considered only as a first step in the
elucidation of their structure.

\section{Strong decays of $a_{0}(1450)$ and $K_{0}^{\ast}(1430)$.}

As discussed above, the introduction of glueball fields in the model is
expected to modify its predictions for the isosinglet sectors. However, the
predictions of the model for isodoublets and isotriplet sectors are not
expected to be substantially modified by the introduction of glueball fields.
Thus it is worthy to work out the predictions of the model for the decay
widths in these sectors. Unfortunately there is no confident experimental
information for pseudoscalar fields \cite{PDG}, hence we will analyze the
model predictions for the scalar fields where more accurate data exists. The
total widths of $a_{0}(1450)$ and $K_{0}^{\ast}(1430)$ are \cite{PDG}%
\begin{equation}
\Gamma\left[  a_{0}(1450)\right]  _{Exp}=265\pm13\text{ MeV,\qquad}%
\Gamma\left[  K_{0}^{\ast}(1430)\right]  _{Exp}=294\pm23\text{ MeV,}%
\end{equation}
from where we obtain the fraction
\begin{equation}
\left.  \frac{\Gamma\left[  a_{0}(1450)\right]  }{\Gamma\left[  K^{\ast
}(1430)\right]  }\right\vert _{Exp}=0.90\pm0.08\text{ }.\label{AK}%
\end{equation}
The measured partial widths are%
\begin{align}
\left.  \frac{\Gamma\left[  a_{0}(1450)\rightarrow\pi\eta^{\prime}\right]
}{\Gamma\left[  a(1450)\rightarrow\pi\eta\right]  }\right\vert _{Exp} &
=0.35\pm0.16,\\
\left.  \frac{\Gamma\left[  a_{0}(1450)\rightarrow K\overline{K}\right]
}{\Gamma\left[  a_{0}(1450)\rightarrow\pi\eta\right]  }\right\vert _{Exp} &
=0.88\pm0.23,\label{AKKApieta}\\
\frac{\Gamma\left[  K_{0}^{\ast}(1430)\rightarrow\pi K\right]  }{\Gamma\left[
K_{0}^{\ast}(1430)\rightarrow\text{all}\right]  } &  =0.93\pm0.04\pm0.09\text{
}.
\end{align}
From the theoretical side these decays have been studied using $SU(3)$
symmetry \cite{twononetsyr} with the following predictions%
\begin{align}
\left.  \frac{\Gamma\left[  a_{0}(1450)\rightarrow\pi\eta^{\prime}\right]
}{\Gamma\left[  a_{0}(1450)\rightarrow\pi\eta\right]  }\right\vert _{SU(3)} &
=0.16,\\
\left.  \frac{\Gamma\left[  a_{0}(1450)\rightarrow K\overline{K}\right]
}{\Gamma\left[  a_{0}(1450)\rightarrow\pi\eta\right]  }\right\vert _{SU(3)} &
=0.55,\\
\left.  \frac{\Gamma\left[  a_{0}(1450)\right]  }{\Gamma\left[  K^{\ast
}(1430)\right]  }\right\vert _{SU(3)} &  =1.51,
\end{align}
which are in disagreement with experimental results.

In our formalism these decays are induced at the three level and are given by
\begin{equation}
\Gamma(S\rightarrow P_{1}P_{2})=\frac{1}{8\pi}\left\vert g_{SP_{1}P_{2}%
}\right\vert ^{2}\frac{\left\vert \mathbf{p}_{1}\right\vert }{m_{s}^{2}}%
\end{equation}
where $g_{SP_{1}P_{2}}$ stands for the corresponding coupling which is
extracted from Eq. (\ref{L3}). Using our results for the couplings in Table IV
we get%
\begin{align}
\Gamma\left[  a_{0}(1450)\rightarrow\pi\eta\right]   &  =116.8\pm57.6\text{
MeV,}\\
\Gamma\left[  a_{0}(1450)\rightarrow\pi\eta^{\prime}\right]   &
=0.00\pm0.03\text{ MeV,}\\
\Gamma\left[  a_{0}(1450)\rightarrow K^{+}K^{-}\right]   &  =76.5\pm45.6\text{
MeV,}%
\end{align}%
\begin{equation}
\Gamma\left[  K^{\ast}(1430)\rightarrow\pi K\right]  =396.6\pm262.4\text{
MeV.}%
\end{equation}
The decay width $\Gamma\left[  a(1450)\rightarrow K\bar{K}\right]  $ =
$2\Gamma\lbrack a(1450)\rightarrow$ $K^{+}K^{-}]$ and assuming those three
decay modes are the dominant ones we get%
\begin{equation}
\Gamma\left[  a_{0}(1450)\right]  =269.9\pm107.9\text{ MeV.}%
\end{equation}
This yields%
\[
\frac{\Gamma(a(1450)\rightarrow K\bar{K})}{\Gamma(a(1450)\rightarrow\pi\eta
)}=1.31\pm1.01
\]
consistent with the value quoted in Eq. (\ref{AKKApieta}) for this fraction.
Notice however that the fraction of the $\pi\eta^{\prime}$ to $\pi\eta$
partial widths turns out to be small in the model as compared with experiment
a result due to the prediction of a decay width consistent with zero for
$\pi\eta^{\prime}$ channel. Again, the $\eta^{\prime}$ is expected to contain
certain amount of glueball which is not included in the present context. For
the fraction in Eq. (\ref{AK}) the model yields%

\begin{equation}
\frac{\Gamma\left[  a(1450)\right]  }{\Gamma\left[  K^{\ast}(1430)\right]
}=0.68\pm0.53
\end{equation}
consistent with experimental results.

\section{Summary and perspectives.}

Summarizing, in this paper we point out the existence of a quasi-degenerate
chiral nonet around 1.4 GeV whose scalar part presents a slightly inverted
spectrum as compared with expectations from a conventional $\overline{q}q$
nonet. Guided by the empirical linear rising of the mass of a hadron with the
number of constituent quarks and the fact that a four-quark nonet should
present an inverted spectrum as pointed out long ago by Jaffe
\cite{Jaffe:1977}, we conjecture that this cuasi-degenerate chiral nonet and
the lowest lying scalar and pseudoscalar mesons, come from the mixing of a
chiral $\overline{q}\overline{q}qq$ nonet (chiral symmetry realized directly)
with a conventional $\overline{q}q$ chiral nonet (chiral symmetry
spontaneously broken). We realize these ideas in the framework of an effective
chiral model. We introduce quark mass terms appropriate to four-quark states
which give to these states an inverted spectrum. Particularly important for
the $\overline{q}q$ states is the instanton induced effective trilinear
interaction. We fix the parameters of the model using as input the masses for
$\pi(137)$, $a_{0}(980)$, $K(495),$ $\eta(548),$ $\eta^{\prime}(957),$
$\eta(1295)$ in addition to the weak decay constants $f_{\pi}$ and $f_{K}$.
The model gives definite predictions for the masses of the remaining members
of the nonets and the couplings of all mesons entering the model. The outcome
of the fit allow us to identify the remaining mesons as the $\pi(1300),$
$\eta(1450),$ $K(1469)$ on the pseudoscalar side and $f_{0}(1370),$
$K_{0}^{\ast}(1430),$ $a_{0}(1450),$ $f_{0}(1500)$ on the scalar side. In
general, isospinor and isovector pseudoscalar mesons arise as weak mixing of
$\overline{q}q$ and tetraquark fields. In contrast, scalars in these isotopic
sectors get strongly mixed. We obtain also strong mixing between $\overline
{q}q$ and tetraquarks for isoscalar fields. We work out the model predictions
for the most relevant couplings and calculate explicitly the strong decays of
the $a_{0}(1450)$ and $K_{0}^{\ast}(1430)$ mesons. From the comparison of the
predicted couplings with the experimental ones we conclude that observable for
the isovector and isospinor sectors are consistently described within the
model. However, although the masses in the isoscalar sectors are consistent
with the conjecture mentioned above, the couplings are not all properly
described. However, the states entering the model lie at an energy quite close
to the expected mass for the lightest glueballs. We expect the inclusion of
these states to provide the missing ingredients to reach a consistent global
picture of pseudoscalar and scalar mesons below $1.5$ $GeV$. The introduction
of these fields in the present context can be easily done since these states
behave as singlets with respect to chiral transformations. Furthermore these
fields can acquire non-vanishing vacuum expectation values rendering the
glueball-meson-meson interactions natural candidates to generate the (somewhat
artificial) mixing term (or similar ones) in Eq. (\ref{Leps}) in a simple and
straightforward way. On the experimental side, there is a unique candidate in
the scalar sector for the missing state, the $f_{0}(1710)$, once we introduce
the lightest scalar glueball. All states above 1 GeV, including the
$f_{0}(1710)$, are expected to acquire a glueball component to some extent.

Another piece of information which could prove relevant is a direct mixing
between $\ \overline{q}q$ and $\overline{q}\overline{q}qq$ states due to
instantons which is not included here. These possibilities are presently under
investigation. Predictions of the model for the two photon decay of neutral
scalars will be published elsewhere \cite{twophotons}.

\begin{acknowledgments}
Work supported by CONACYT M\'{e}xico under project 37234-E.
\end{acknowledgments}

\end{document}